\pdfoutput=1

\documentclass[aps,pre,reprint,superscriptaddress]{revtex4-1}

\usepackage{graphicx}
\usepackage{amsmath}
\usepackage{amssymb}
\usepackage{caption}
\usepackage{subcaption}

\DeclareMathAlphabet{\mathbbold}{U}{bbold}{m}{n} %for unity operator: use \mathbbold{1} in math area; can't use bbold as package because this would change the amssymb that would also be called with \mathbb 

\newcommand*\PCa{\includegraphics[height=2ex]{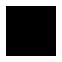}}
\newcommand*\PCb{\includegraphics[height=2ex]{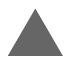}}
\newcommand*\PCc{\includegraphics[height=2ex]{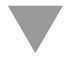}}
\newcommand*\PCd{\includegraphics[height=2ex]{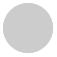}}

\begin{document}

\title{Modulational instability and resonant wave modes act on the metastability of oscillator chains}
\author{Torsten Gross}
\affiliation{Department of Physics, Humboldt-Universit\"at zu Berlin, D-12489 Berlin, Germany}
\email{tgross@physik.hu-berlin.de}  

\author{Dirk Hennig}
\affiliation{Department of Mathematics, University of Portsmouth, Portsmouth PO1 3HF, United Kingdom}
\email{dirk.hennig@port.ac.uk}  

\author{Lutz Schimansky-Geier}
\affiliation{Department of Physics, Humboldt-Universit\"at zu Berlin, D-12489 Berlin, Germany}
\email{alsg@physik.hu-berlin.de}

\date{June 4, 2014}

\begin{abstract}
We describe the emergence and interactions of breather modes and resonant wave modes within a two-dimensional ring-like oscillator chain in a microcanonical situation. Our analytical results identify different dynamical regimes characterized by the potential dominance of either type of mode. The chain is initially placed in a meta-stable state which it can leave by passing over the brim of the applied Mexican-hat-like potential. 
%This escape process can be realized through two different escape channels which are defined by different transition states. 
We elucidate the influence of the different wave modes %and escape channels 
on the mean-first passage time. A central finding is that also in this complex potential landscape a fast noise-free escape scenario solely relying on nonlinear cooperative effects is accomplishable even in a low energy setting.

\end{abstract}

\maketitle

\section{Introduction}

The interest in the escape of coupled degrees of freedom or chains of interacting units out of metastable states has been intensified lately \cite{Sung, Park, Dikshtein, Sebastian, Lee, Kraikivsky, EPL, escape_world_scientific}. Escape is accomplished when the considered object overcomes a potential barrier separating the local minimum of the potential landscape from a neighbouring domain of attraction. The point of lowest energy along this barrier is a saddle point in the potential landscape and its related configuration is referred to as a transition state. This so called activation energy is therefore minimally required to surmount the energetic bottleneck and can be provided in two different ways. One is the possibility of stochastic escape occurring in the presence of a heat bath that is sampled for the optimal fluctuations triggering an event of escape \cite{Langner,Hanggi}. The second one is that in the noise free situation the energy can be supplied in a single shot under micro-canonical circumstances. 

Previous work \cite{escape_world_scientific, EPL} addressed microcanonical escape scenarios of chains of interacting units in one- and two-dimensional potential landscapes with a single barrier and compared those to the corresponding noise-assisted escape process. In the deterministic setting a nonlinear breather dynamics assists at a speedy passage through the metastable transition state. If additionally noise and linear dissipation acts on the chain and the friction is large, the escape rate becomes smaller. The breathers are then unable to grow out of the phonon background. In contrast for small friction when relaxation times become large, the breathers are able to survive and contribute as added noise to the amplification of the escape \cite{escape_world_scientific}.

We wish to extend these studies to systems of higher complexity and explore whether the observed phenomena are solely inherent to highly idealized settings or whether they remain relevant in a more general context. To this end, we investigate a ring of interacting units evolving in a Mexican-hat-like two-dimensional potential landscape. A detailed inspection of the chain dynamics yields an understanding of the emergence of breathers and resonant wave modes. This leads to analytical results that allow to determine parameter choices which accomplish an efficient noise-free escape scenario solely relying on nonlinear cooperative effects. 

Apart from our conceptual interest in this work, the results can be applied to the description of micro bubble surface modes. Micro bubble surfaces can be modelled by a closely related system in which breather modes were verified experimentally \cite{micro_bubbles}. Our results demonstrate the potential relevance of resonant wave modes and the escape behaviour within this specific context.

The paper is organised as follows: In the next section the model of the ring of interacting units evolving in the Mexican-hat-like potential is introduced. Then, the formation of breather solutions initiated by modulational instability is considered, followed by the analysis of the resonant longitudinal wave modes and the interaction between the two. After establishing this theoretical framework, the subsequent section elucidates the deterministic escape scenario. Transition states and the associated escape channels are discussed in detail, followed by an investigation of the escape time statistics. Finally, we summarise our results.

\section{Localised and resonant wave modes in an oscillator chain model\label{sec:waves}}

We study a Hamiltonian system consisting of a two dimensional chain of $N$ linearly coupled oscillators of mass $m$ subjected to an external Mexican-hat-like an-harmonic potential $V(\mathbf{q}_i)=-a\, \sqrt{\mathbf{q}_i^2}+b\,\cos\left( \sqrt{\mathbf{q}_i^2} / \lambda\right)$. The chain shall be closed so that the 2D coordinates and momenta are subject to periodic boundary conditions, $\mathbf{q}_{N-1}=\mathbf{q}_0\quad \mbox{and}\quad \mathbf{p}_{N-1}=\mathbf{p}_0.$ Rescaling coordinates, $\mathbf{\widetilde{q}}_i=(b/a)\,\mathbf{q}_i$, momenta, $\mathbf{\widetilde{p}}_i=\sqrt{m\,b}\;\mathbf{p}_i$, and time, $\tilde{t}=(\sqrt{m\,b}/a)\,t$, to natural units yields two remaining effective parameters, the coupling strength $\widetilde{\kappa}=(b/a^2)\,\kappa$ and the potential width parameter $\widetilde{\lambda}=(a/b)\,\lambda$, and leads to the Hamiltonian (tildes have been omitted)
\begin{align}\label{scaledHam}\mathcal{H}&=\sum_{i=0}^{N-1}\left[\frac{\mathbf{p}_i^2}{2}+\frac{\kappa}{2}\left(\mathbf{q}_i-\mathbf{q}_{i+1}\right)^2+V\left(\mathbf{q}_i\right)\right],\\ \nonumber
V\left(\mathbf{q}_i\right)&=-\sqrt{\mathbf{q}_i^2}+\cos{\left( \frac{\sqrt{\mathbf{q}_i^2}}{\lambda}\right) },
\end{align}
with the corresponding equations of motion
\begin{align}
\begin{split}\label{qdyn1}
\mathbf{\ddot{q}}_i=&-\kappa\,\left(2\,\mathbf{q}_i-\mathbf{q}_{i+1}-\mathbf{q}_{i-1}\right)+\frac{\mathbf{q}_i}{\sqrt{\mathbf{q}_i^2}}\\
&+\sin{\left(\frac{\sqrt{\mathbf{q}_i^2}}{\lambda}\right)}\,\frac{\mathbf{q}_i}{\lambda\,\sqrt{\mathbf{q}_i^2}}\qquad i\in0\ldots N-1.
\end{split}
\end{align}
The energy is conserved
\begin{align*}
\mathcal{H}\left(\left\lbrace\mathbf{p}(t)\right\rbrace,\left\lbrace\mathbf{q}(t)\right\rbrace\right)=\text{const.}=E.
\end{align*}

Considering the rotational symmetry of the Mexican-hat-like potential, it is convenient to  express the equations of motion also in terms of radial and angular coordinates, $r_i$ and $\varphi_i$,
\begin{align}\begin{split}\label{rdyn1}
\ddot{r}_i=&r_i\,\dot{\varphi}_i^2-\kappa\,(2\,r_i-r_{i+1}\,\cos{\left(\varphi_i-\varphi_{i+1}\right)}\\
&-r_{i-1}\,\cos{\left(\varphi_{i-1}-\varphi_i\right)})+1+\frac{1}{\lambda}\,\sin{\left(\frac{r_i}{\lambda}\right)}
\end{split}\\
\begin{split} \label{phidyn1}
\ddot{\varphi}_i\,r_i=&-2\,\dot{r}_i\,\dot{\varphi}_i+\kappa(r_{i-1}\,\sin\left(\varphi_{i-1}-\varphi_{i}\right)\\
&-r_{i+1}\,\sin\left(\varphi_{i}-\varphi_{i+1}\right)).
\end{split}
\end{align}

\begin{figure}
\centering
\includegraphics[width=\linewidth]{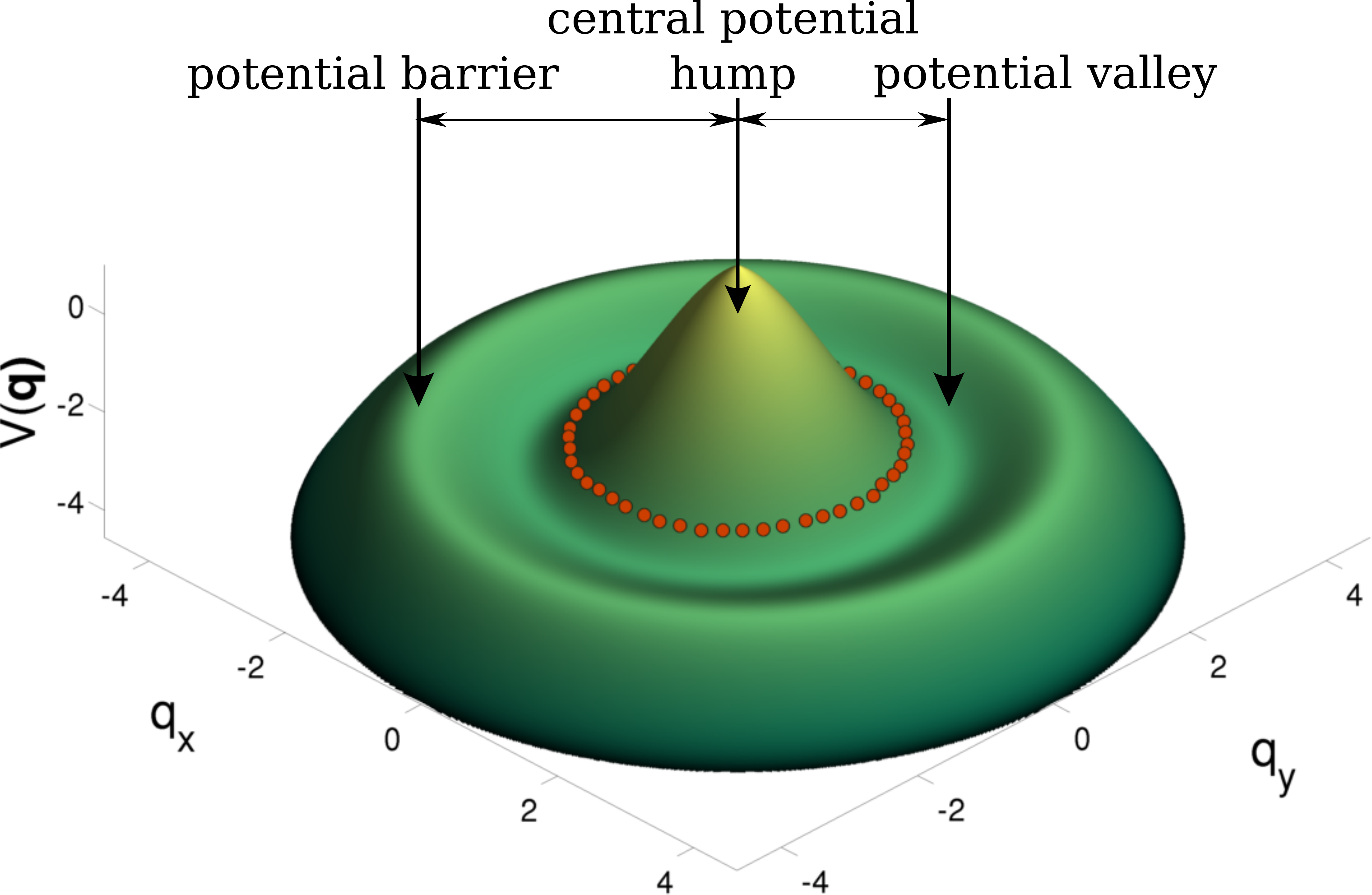}
\caption{A typical initial preparation of the oscillator chain.}
\label{fig:IC}
\end{figure}

\subsection*{Stability of the minimum energy configuration}

Fixed points are found for the configuration $r_i=r^0$, $\varphi_i=i\,\Delta \Theta$, with $\Delta \Theta=2\pi/N$, where a conditional equation for $r^0$ arises from setting all time derivatives in Eq. \eqref{rdyn1} to zero
\begin{align}\label{r^0-conditional_equation}
0=-2\,\kappa\,r^0\left(1-\cos\left(\frac{2\,\pi}{N}\right)\right)+1+\frac{1}{\lambda}\sin\left(\frac{r^0}{\lambda}\right).
\end{align}
Initially, the chain is placed in the vicinity of this so called minimum energy configuration. In order for the chain to evolve from a metastable state, that is from a basin of potential energy, the fixed point needs to be (Lyapunov) stable. Consequently, we will need to limit our parameter space to the region of stability. The results of a linear stability analysis, see Appendix \ref{App:stabl_fp}, are represented in Fig.\,\ref{fig:min_E_config_stability}. It shows instability arising for large values of the coupling constant when the chain's tendency to contract becomes so strong that any small perturbation will initiate a contraction that pulls the entire chain over the central potential hump into the first potential valley (and possibly beyond). 

As an additional constraint on the parameters, let us require the existence of the barrier in the Mexican-hat-like potential (see Fig.\,\ref{fig:IC}), which confines the range of the potential width parameter to $0 < \lambda < 1$ 
and  in conjunction with  a restriction of  the coupling constant to $0< \Delta\Theta ^2 \kappa < 1.3$  a sufficiently large region of stability in parameter space results. The number of oscillators will be kept at $N=100$.

\begin{figure}
\centering
  \includegraphics[height=4.05cm]{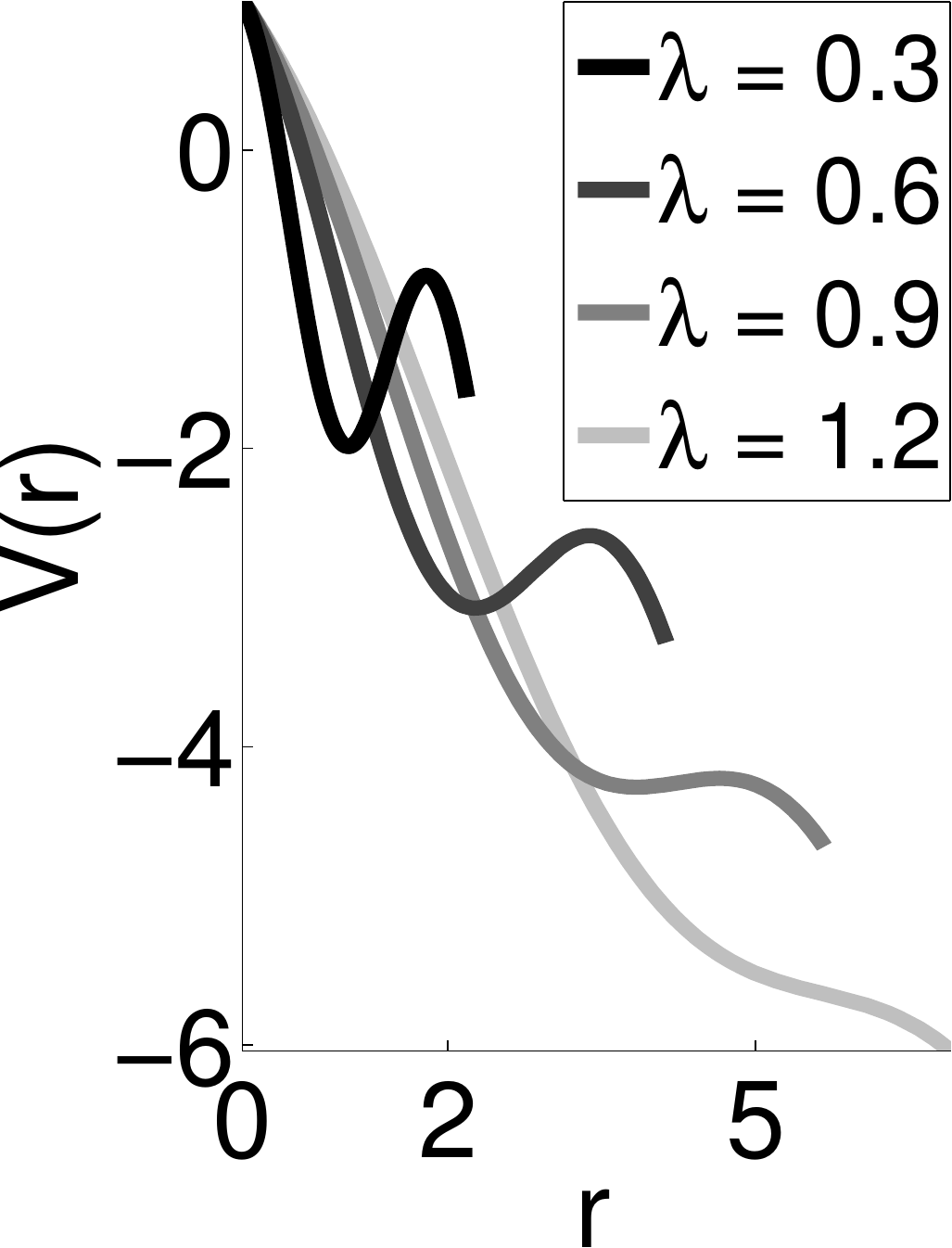}
  \hfill
  \includegraphics[height=4.2cm]{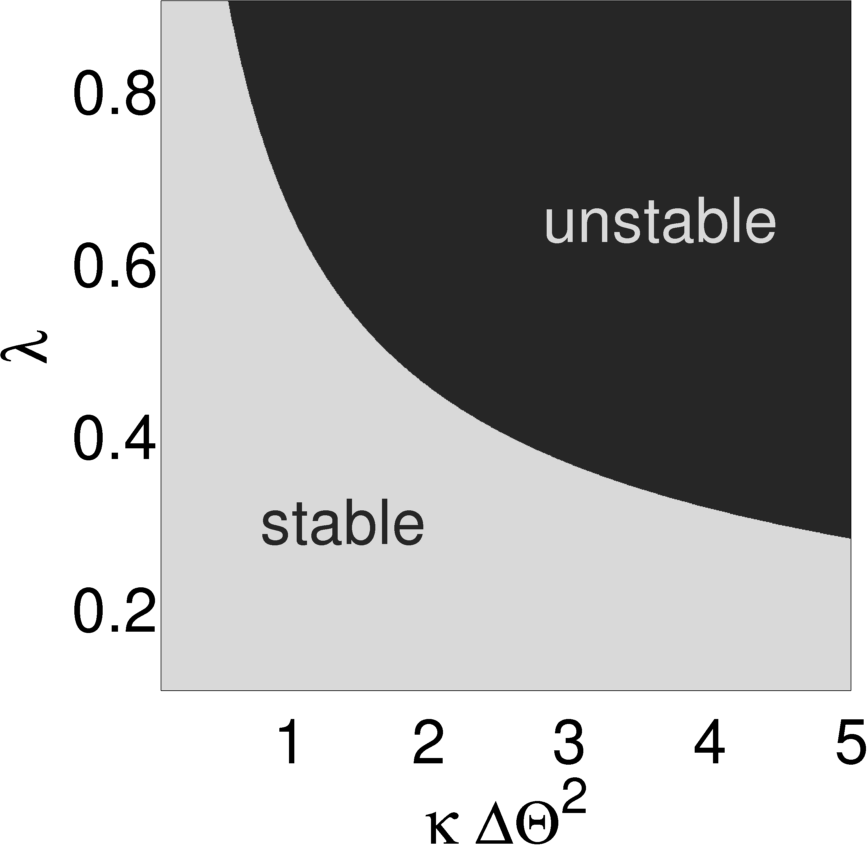} 
    \caption{The left figure represents the profile of the Mexican-hat-like potential. The potential barrier vanishes for $\lambda>1$. The right figure shows the result of the stability analysis of the fixed point comprising the minimum energy configuration and vanishing momenta.}
\label{fig:min_E_config_stability}
\end{figure}

\subsection*{Modulational instability produces radial breathers}

Initially, the oscillators are placed in a perturbed ring-like structure around the central potential hump (Fig.\,\ref{fig:IC}),
\begin{align*}
\mathbf{q}_i^{\mbox{IC}}=r^{\mbox{IC}}\cdot\begin{pmatrix}\cos\left(i\,\Delta \Theta \right)\\ \sin\left(i\,\Delta \Theta \right)\end{pmatrix}+\begin{pmatrix}
\Delta q_i^x\\ \Delta q_i^y
\end{pmatrix},\qquad
\mathbf{p}_i^{\mbox{IC}}=\begin{pmatrix}
\Delta p_i^x\\ \Delta p_i^y
\end{pmatrix},
\end{align*}
where $\Delta q_i^x$ and $\Delta q_i^y$ as well as $\Delta p_i^x$ and $\Delta p_i^y$ are random perturbations taken from a uniform distribution within the intervals
\begin{align*}
\Delta q_i^x,\,\Delta q_i^y\in \left[-0.01,0.01 \right];\quad
\Delta p_i^x,\,\Delta p_i^y\in \left[-0.01,0.01 \right].
\end{align*}
In this setting, the angular distance between any pair of neighbouring oscillators is almost equal (close to $\Delta \Theta$) so that the initial angular acceleration is small. Thus, for short time periods after the system's preparation, the assumption $\varphi_i(t)=\varphi_i^0=i\, \Delta \Theta$ is plausible. Likewise, for low energy settings each oscillator's initial radius is close to $r^0$, so that the equation of motion for the radial components can be approximated by a Taylor expansion in $r_i$ around $r_i={r_i}^0$, where we will neglect terms of order ${\Delta r_i}^3=\left(r_i-r^0\right)^3$ and higher. Thus, as an approximation for short periods of time after the initiation of the system, the angular components remain fixed and the evolution of the radial components is governed by the equation
\begin{align*}\begin{split}
\Delta \ddot{r}_i&=\kappa\left(\cos\left(\frac{2\,\pi}{N}\right)\left(\Delta r_{i+1}+\Delta r_{i-1}\right)-2\,\Delta r_i\right)\\
&-{\omega _0}^2\,\Delta r_i+\alpha\, \Delta {r_i}^2,
\end{split}\\ \nonumber
{\omega _0}^2&=-\frac{1}{\lambda ^2}\cos\left(\frac{r^0}{\lambda}\right),\qquad \alpha=-\frac{1}{2\,\lambda ^3}\sin\left(\frac{r^0}{\lambda}\right).
\end{align*}
In this case, the oscillators can only move along equally spaced rays that all emerge from the origin. Therefore this approximation will be referred to as the star-like chain model.
\begin{figure}
\centering
\parbox{0.32\linewidth}{time: 111.8}\hfill
\parbox{0.32\linewidth}{time: 173.2}\hfill
\parbox{0.32\linewidth}{time: 201.8}\\ 
\vspace{1ex}
\includegraphics[width=0.32\linewidth]{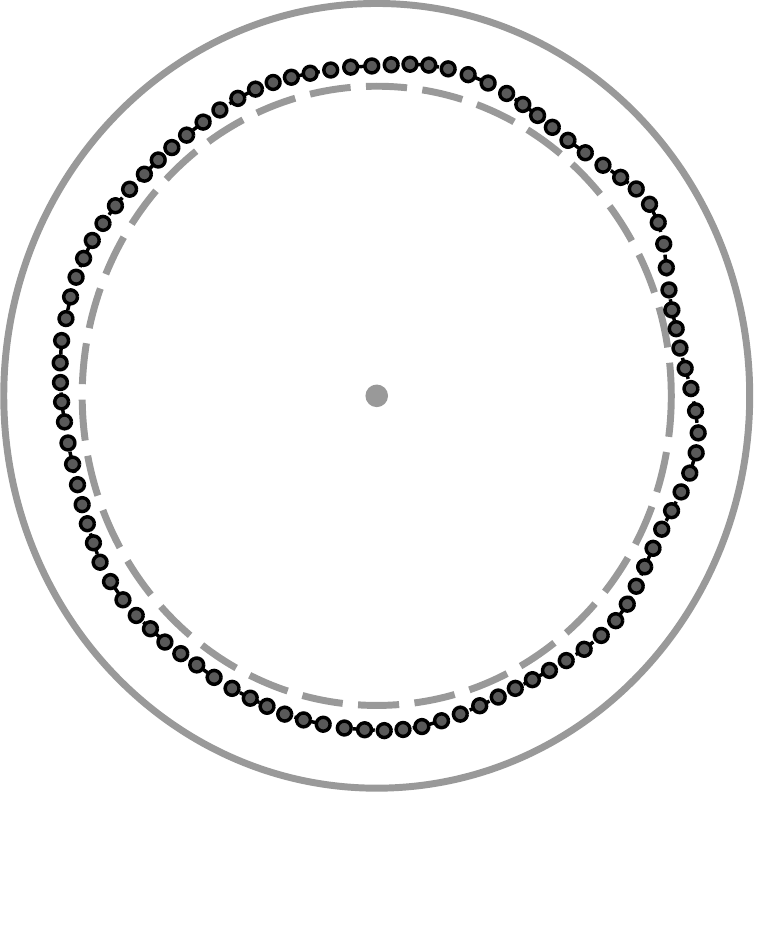}\hfill
\includegraphics[width=0.32\linewidth]{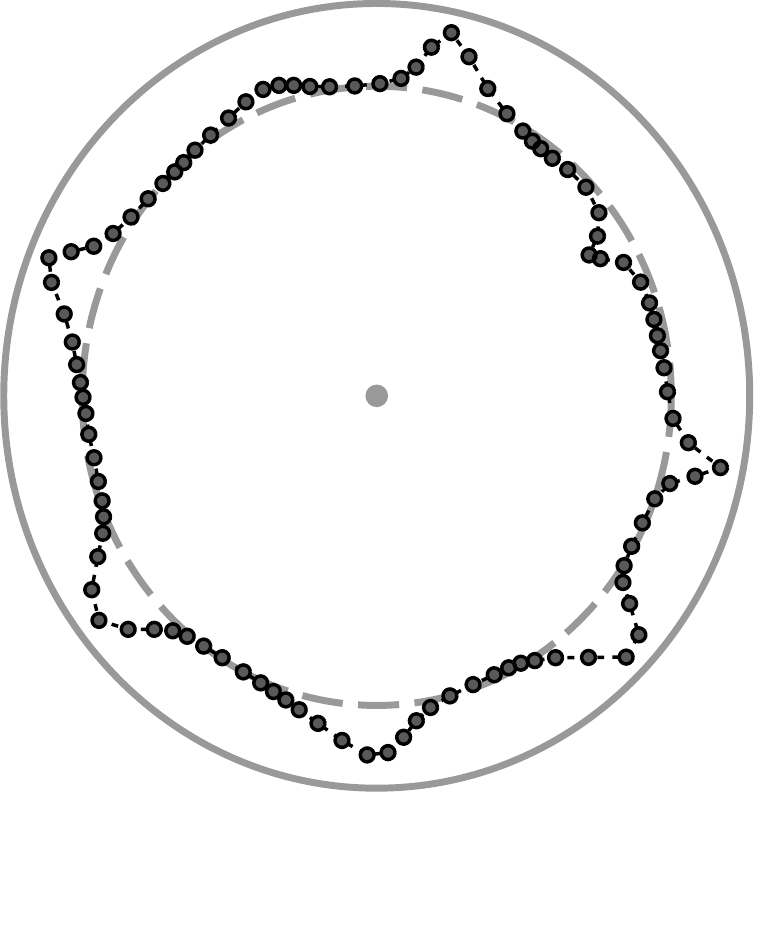}\hfill
\includegraphics[width=0.32\linewidth]{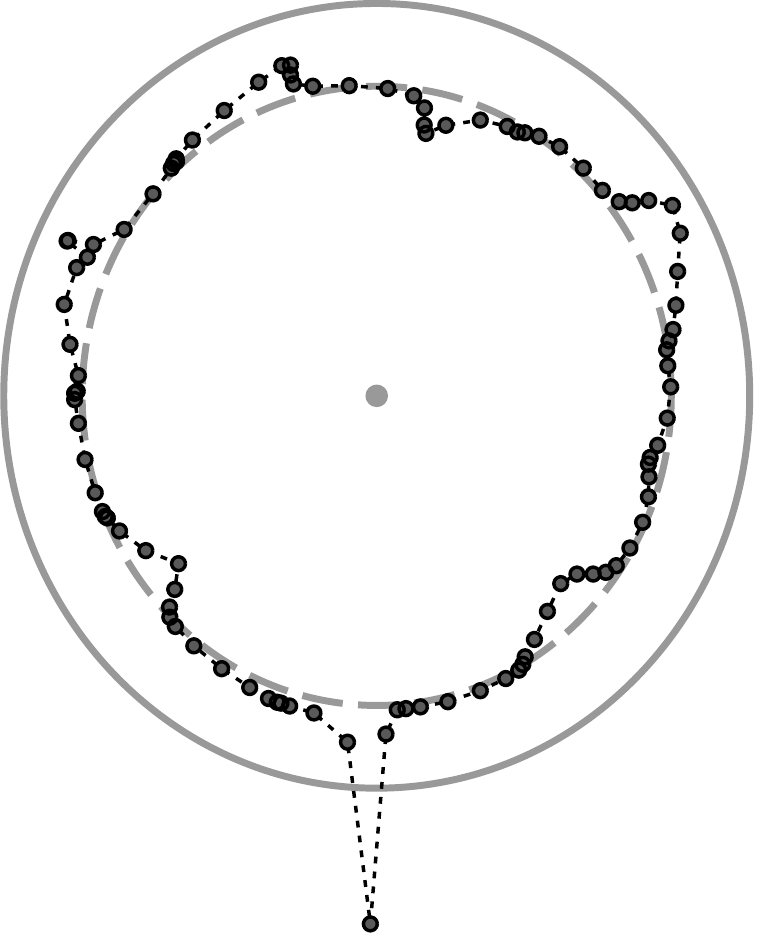}
\caption{Simulation snapshots showing the growth of a radial breather array from an almost homogeneous initial state due to modulational instability. The ongoing amplification of this pattern eventually drives an individual oscillator over the potential barrier and thus triggers an escape of type I as described in Sect \ref{sect:det-esc}. Parameters: $\kappa\,\Delta\Theta ^2=0.79\cdot 10^{-4}$, $\lambda=0.85$.}
\label{fig:MI_high_wavenumber_escape}
\end{figure}

Due to the initial ring-like set-up the chain will first oscillate in a $k=0$ phonon-like manner. Typically, in our simulations we observe  the emergence of a regularly spaced array of breathers, as shown in Fig.\,\ref{fig:MI_high_wavenumber_escape}. These localised excitations play a crucial role because they concentrate energy in single degrees of freedom and therefore substantially influence the escape behaviour, as will be discussed in \ref{sect:det-esc}.

The studies in \cite{MI_1,MI_2} reveal that the star-like chain model is able to describe the emergence of these transversal modes in terms of a modulational instability. The key idea herein is to formulate a discrete nonlinear Schr\"odinger equation (DNLS), also known as the discrete self-trapping equation \cite{DST}, for a time dependent amplitude of the first-harmonic phonon wave. Inspecting the stability of the DNLS's plane wave solutions yields a dispersion relation for the small perturbation terms which will allow to determine the wave modes on the oscillator chain that grow in amplitude and create the transversal pattern. We call the wave number and the angular frequency of the phonons $k$ and $\omega$, and those of the perturbations $Q$ and $\Omega$. Their dispersion relations read
\begin{gather}\nonumber
\omega ^2={\omega _0}^2+2\kappa\left(1-\cos\Delta \Theta\,\cos k\right) \quad \text{and}\\
\label{pert.-disp_rel}
\begin{split}\left(\omega_0\, \Omega-\kappa\cos\Delta\Theta\,\sin k\,\sin Q\right)^2=
\kappa\cos\Delta\Theta\,\cos k \\ \times \sin ^2 \frac{Q}{2}\left(4\kappa\,\cos \Delta\Theta \,\cos k \, \sin ^ 2 \frac{Q}{2}-2\gamma A^2 \right).
\end{split}
\end{gather}
Eq. \eqref{pert.-disp_rel} describes the stability of the $Q$-mode perturbation on the $k$-mode carrier wave. $Q$ and $k$ have a $2\,\pi$ periodicity and can therefore be chosen to be in the first Brillouin zone. Furthermore we can restrict the range of $k$ and $Q$: $k,Q\in [ 0 ,\pi ]$, because negative values only correspond to waves with the opposite direction of propagation. The perturbations are stable for $\Omega\in\mathbb{R}$ which is the case when the right hand side of \eqref{pert.-disp_rel} is positive. Since $\gamma\geq 0$ all carrier waves with $\cos k\leq 0 \Leftrightarrow k\in [\pi/2, \pi ]$ are therefore stable with respect to any perturbation mode. For $k\in [ 0 , \pi/2 ]$ perturbations will grow, provided that
\begin{align}\label{MI-unstable_Q}
\cos k \, \sin ^2\left(\frac{Q}{2}\right)&\leq \frac{A^2}{{A_0}^2}.
\end{align}
Here $A$ is the phonon amplitude,
\begin{align*}
{A_0}^2&=\frac{2\kappa\cos(\Delta\Theta)}{\gamma},\quad \text{and} \quad \gamma = \frac{10\alpha ^2}{3{\omega _0}^2}.
\end{align*} 
We can then find an according growth rate
\begin{multline}\label{MI-growth_rate}
\widetilde{\Gamma}_r (Q)=\left |\operatorname{Im} (\Omega) \right |=\frac{\sin\left(\frac{Q}{2}\right)}{\omega _0}\\
\times \sqrt{2\,\kappa\,\gamma\,\cos k\,\left({A}^2-{A_0}^2 \sin^2 \left(\frac{Q}{2}\right)\cos k\right)}.
\end{multline}
The left-hand side of \eqref{MI-unstable_Q} is monotonically increasing within the range of $Q$ and therefore defines an upper bound, $Q^*$, above which wave modes are no longer unstable. 
For the case
$
A^2\leq 2\, {A_0}^2\,\cos k,
$ function \eqref{MI-growth_rate} attains its maximum at
\begin{equation*}%\label{MI-Q_max}
Q_{max}=2\,\arcsin\sqrt{\frac{A^2}{2\,{A_0}^2\, \cos k}}.
\end{equation*}
If this is not an unstable wave mode because $Q_{max}>Q^*$, this upper bound $Q^*$ become the most unstable mode because \eqref{MI-growth_rate} is monotonically increasing for $Q\in [ 0 ,Q_{max} ]$.
For $A^2> 2\, {A_0}^2\,\cos k$, the maximum growth rate is found at $Q=\pi$. At the maximum of the growth rates the most unstable mode emerges characterising the spatial pattern and temporal growth rate of the breather array. 

For our simulations the system is prepared in a flat initial state with small random perturbations, we can therefore set $k=0$. This also means that the phonon wave amplitude $A$ is related to the energy of the system, $E$, through
\begin{align}
\label{eq:isolate_A}
E=\mathcal{H}\left(\left\lbrace \mathbf{p}_i(t)=0\right\rbrace ,\left\lbrace r_i=r^0+A,\varphi_i={\varphi_i}^0\right\rbrace \right).
\end{align}
Thus, isolating $r^0$ from Eq. \eqref{r^0-conditional_equation} allows to calculate $A$ from the above equation for given energy and system parameters and therefore to determine mode number and growth rate of the most unstable modulation. However, due to the periodic boundary conditions only a discrete spectrum of perturbative waves can occur
\[Q=m_r\,\Delta \Theta \quad m_r\in0,1\ldots (N-1), \]
where $m_r$ denotes the possible mode numbers. To take account of this, we choose the wave number of the most unstable (continuous) mode as the nearest value lying in the corresponding unstable part of the discrete spectrum yielding the predicted mode number of the emerging transversal wave. Its dependence on the system parameters is represented in Fig.\,\ref{fig:MI_m_max} together with the according growth rates in Fig.\,\ref{fig:MI_Gamma_max}.

\subsection*{Emergence of resonant longitudinal waves}

\begin{figure}
\centering
\parbox{0.32\linewidth}{time: 1.1}\hfill
\parbox{0.32\linewidth}{time: 37.2}\hfill
\parbox{0.32\linewidth}{time: 40.15}\\ 
\vspace{1ex}
\includegraphics[width=0.32\linewidth]{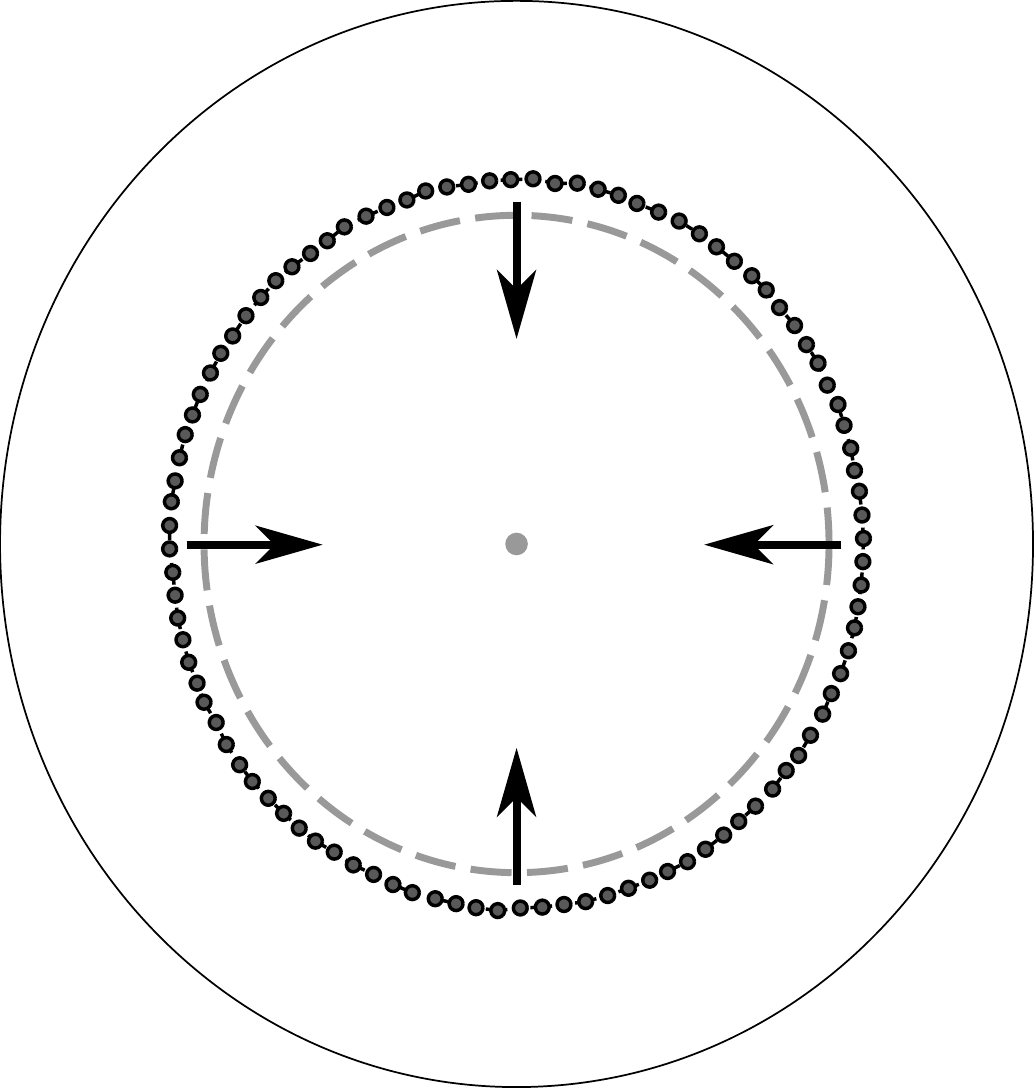}\hfill
\includegraphics[width=0.32\linewidth]{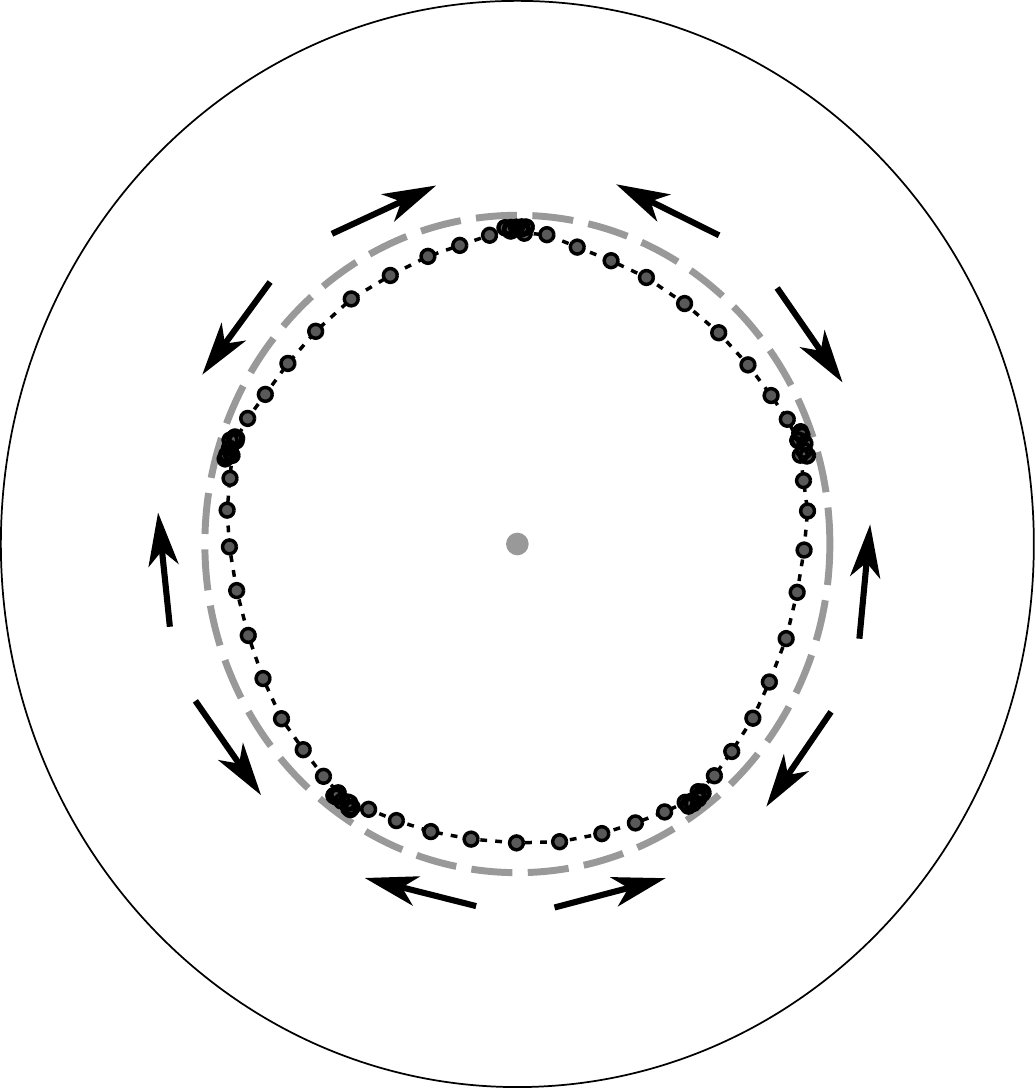}\hfill
\includegraphics[width=0.32\linewidth]{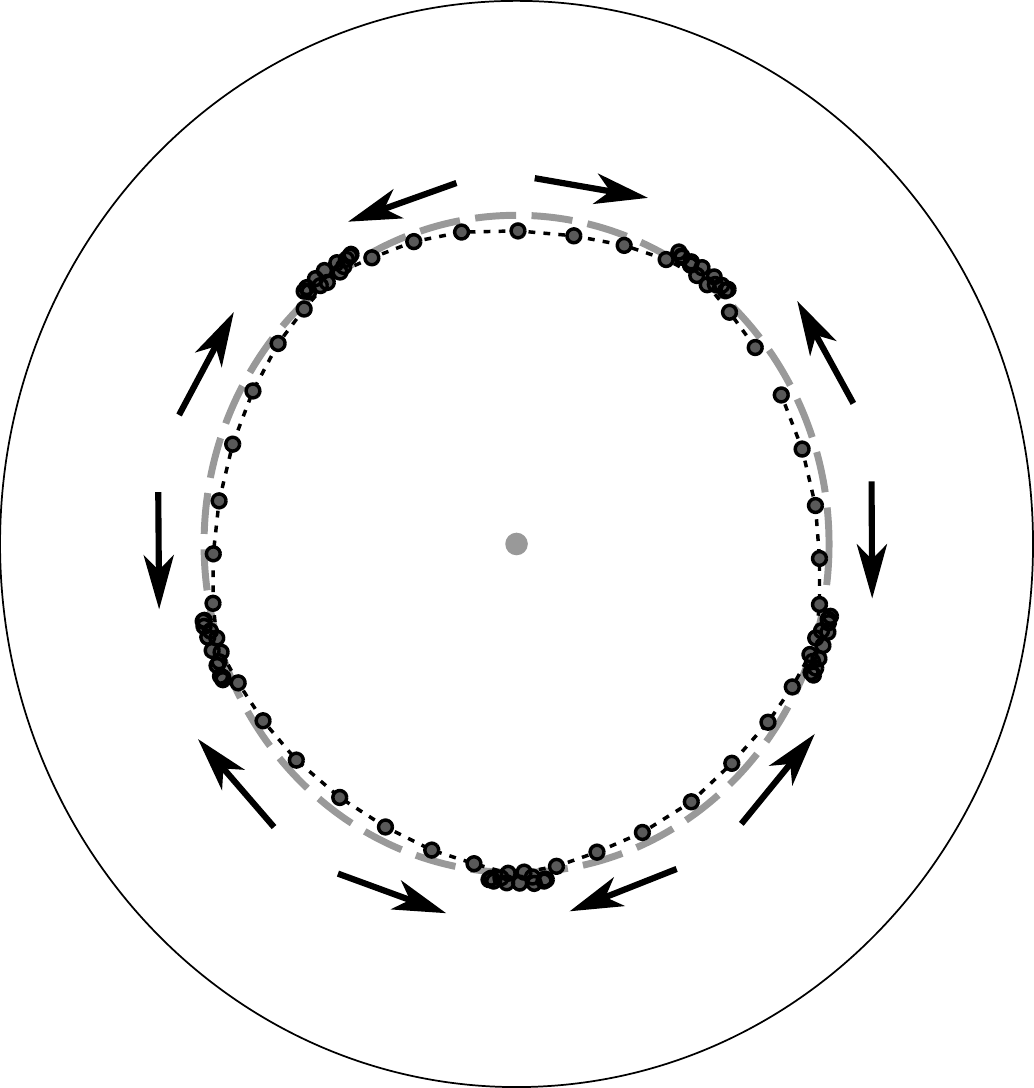}
\caption{Simulation snapshots showing the emergence of a longitudinal wave pattern with wave number $m_{\varphi}=5$. Arrows indicate the chain movement. Parameters: $\kappa\,\Delta\Theta ^2=0.06$, $\lambda=0.4$}
\label{fig:longi_reso_example}
\end{figure}

The foregoing analysis characterises the transversal wave pattern that arises from modulational instability on a $k=0$ phonon background within the star-like chain model. This is an essential step in the understanding of the chain escape behaviour, as discussed in Sect. \ref{sect:transition_state}. However, the basic assumption of fixed angular components is only maintainable for short time periods. Obviously, the angular dynamics plays an increasingly important role as time proceeds. Remarkably, for most of the parameter choices the angular movement is far from being erratic but instead consists of regular and pronounced longitudinal wave patterns, as shown in Fig.\,\ref{fig:longi_reso_example}. These patterns are fundamental for the characterisation of the system's dynamics and its deterministic escape behaviour.

Our analytic description of this phenomenon is motivated by the observation of periodic energy transfer between radial and longitudinal degrees of freedom when these modes appear, see the middle plot of Fig.\,\ref{fig:energy_transf}. Typically, we first observe a few oscillations of the perturbed radial $k=0$ phonon-like mode until the longitudinal pattern grows in amplitude. This causes a decrease of the radial mode's amplitude. After a short time also the longitudinal amplitude decreases again. Thereby the energy is transferred back into the radial degrees of freedom and the entire process repeats. 

Motivated by this observation we expect longitudinal waves to emerge in the presence of the $k=0$ radial phonon-like mode. This leads to the ansatz
\begin{align}\label{para_reso:radial_assumption}
r_i=r^0+A\,\sin (\omega_0\,t)\equiv \widetilde{r}\,^0(t),
\end{align}
with $A<r^0$. Note that this is not an application of (holonomic) constraints which would lead to a new set of equations of motion. Instead Eq. \eqref{para_reso:radial_assumption} serves as an approximation of the original system's radial dynamics. 

It becomes the more inaccurate the more energy is transferred into the angular degrees of freedom (as this reduces the amplitude of the $k=0$ phonon) and the more the initial perturbations cause a deviation from a flat state. However, the aim is to describe a dynamical phenomenon that emerges from a setting where the error of the assumption is small.

Substituting the ansatz \eqref{para_reso:radial_assumption} into the angular equations of motion
\eqref{phidyn1} yields
\begin{align*}
\ddot{\varphi}_i=-\gamma_\varphi(t)\,\dot{\varphi}_i+\kappa\left(\sin\left(\varphi_{i-1}-\varphi_{i}\right)- \sin\left(\varphi_{i}-\varphi_{i+1}\right)\right),
\end{align*}
with $\gamma_\varphi(t)=2\,\dot{\widetilde{r}}\,^0(t)/\widetilde{r}\,^0(t)$.
We can define a continuous angular coordinate $\varphi(\Theta,t)$, where the oscillator index is replaced by a continuous parameter $\Theta\in[0, 2\pi]$. In the continuum limit, $N\rightarrow \infty$, the discrete oscillator index is replaced by the continuous parametrisation variable $\Theta$.  Then, $\varphi(\Theta,t)$ describes the angular deviation of a respective chain segment from the angle $\Theta$ at time $t$, so that we can write $\varphi _i =\left. \varphi(\Theta,t)\right|_{\Theta=i\,\Delta \Theta}+i\,\Delta \Theta$.
Replacing the discrete Laplacian by a continuous second order partial derivative, we arrive at
\begin{align}\label{para_reso:pde}
\ddot{\varphi}(\Theta,t)=-\,\gamma_\varphi(t)\,\dot{\varphi}(\Theta,t)+\kappa\,\left(\Delta\Theta\right)^2\,\frac{\partial^2 \varphi(\Theta,t)}{\partial \Theta^2}.
\end{align}
We can now try to solve it through a separation of variables $\varphi(\Theta,t)=\Phi(\Theta)T(t)$, leading to
\begin{align*}
\frac{\ddot{T}}{T}+\gamma_\varphi\,\frac{\dot{T}}{T}=\kappa\left(\Delta\Theta\right)^2\frac{\Phi^{\prime  \prime}}{\Phi}=\mathrm{const.}\equiv -c^2.
\end{align*}
The $\Theta$-depending part has solutions of the form
\[
\Phi(\Theta)=\Phi_a^0\,\sin\left(\frac{c}{\sqrt{\kappa}\,\Delta\Theta}\,\Theta\right)+\Phi_b^0\,\cos\left(\frac{c}{\sqrt{\kappa}\,\Delta\Theta}\,\Theta\right),
\]
where $\Phi_a^0$ and $\Phi_b^0$ are constants of integration determined by the initial conditions. The periodic boundary conditions (closed chain) imply that $\Phi(\Theta)=\Phi(\Theta+2\pi)$, which restricts the possible values of $c$ to
\begin{align*}%\label{para_reso:c_and_m}
c=m_{\varphi}\,\sqrt{\kappa}\,\Delta\Theta,\quad m_{\varphi}\in\mathbb{N}.
\end{align*}
Along our line of reasoning, the value of $m_{\varphi}$ determines the form of the longitudinal wave pattern.  We recall that $\Phi(\Theta,t)$ quantifies the angular shift between the angle $\Theta$ and the angular coordinate of a chain segment assigned to $\Theta$. Following alongside the course of the chain, the oscillating behaviour of $\Phi$ suggests that we find the chain alternatingly stretched and compressed, compared to the  oscillator spacing in the minimum energy configuration. The frequency of these variations increases with  $m_{\varphi}$. Figure \ref{fig:longi_reso_example} shows the emergence of five stretched and five compressed chain sections each one corresponding to a wave node of $\Phi(\Theta)$. Hence, this longitudinal wave has $m=5$. 

However, certain values of $m_{\varphi}$ represent solutions of Eq. \eqref{para_reso:pde} that are incompatible with the full system \eqref{qdyn1}. First, the Hamiltonian (\ref{scaledHam}) is symmetric with respect to rotation around the origin. Therefore 
the total angular momentum $\mathbf{L}$ is conserved. A longitudinal wave mode with $m_{\varphi}=0$ violates this symmetry, as (see Sect. \ref{A:total_momenta})
\[
|\dot{\mathbf{L}}|=0 \Leftrightarrow \int_0^{2\pi} \Phi(\Theta)\, d\Theta=0 \Leftrightarrow m_{\varphi} \neq 0,
\]
therefore this value will be excluded. Similarly, due to the symmetry of the initial $k=0$ phonon-like mode the total momentum $\mathbf{P}=\lbrace P^x,P^y\rbrace$ is conserved, up to corrections of the order of the random initial perturbations, as long as the chain remains in this setting. It does so from its preparation until the time of the onset of the emergence of either radial or longitudinal modes which we define as $T_{\mathrm{init}} = \min\left(\Gamma_\varphi^{-1},\; \Gamma_r^{-1}\right) $, with $\Gamma_\varphi$ and $\Gamma_r$ being the growth rates of the emerging longitudinal and radial wave modes, see Fig. \ref{fig:M_Gamma_results}. We do not expect longitudinal modes to break this principle. From the results in Sect. \ref{A:total_momenta},
\begin{align*}
\dot{P}^{x/y}(t \ll T_{\mathrm{init}}) \propto \int_0^{2\pi} \sin\Theta\,\sin(m_{\varphi} \,\Theta) = 0 \Leftrightarrow m_{\varphi} \neq 1,
\end{align*}
we can therefore exclude $m_{\varphi}=1$.

Finally, Eq. (\ref{para_reso:pde}) was derived from a continuum limit. However the original system has a finite number of oscillators. The boundary condition thus restricts the upper bound of $m_{\varphi}$ to $N$ as the number of oscillators. In summary, one has
\begin{align*}%\label{para_reso:m_restrict}
m_{\varphi}=2,\,3,\ldots N.
\end{align*}

In relation to Eq. \eqref{para_reso:pde}, the time-dependent part can be interpreted as a parametric oscillator and its term containing the first time derivative cancels out upon a transformation of variables
\begin{align*}\begin{split}
\tau(t)\equiv T(t)\,\exp \int\limits_{0}^{t}  \frac{\gamma_\varphi\left(\,\widetilde{t}\,\right)}{2}\, d\widetilde{t}
= T(t)\,\left(1+\frac{A}{r^0}\sin(\omega_0\,t)\right),
\end{split}
\end{align*}
which leads to a Hill equation
\begin{align}
\label{para_reso:hill-eq}
\ddot{\tau}+c^2\left(1-\frac{\ddot{\widetilde{r}}\,^0(t)}{c^2\,\widetilde{r}\,^0(t)}\right)\tau =0,
\end{align}
where $c$ is called the natural frequency and $-\ddot{\widetilde{r}}\,^0(t)\,/\,\left(c^2\,\widetilde{r}\,^0(t)\right)$ is called pumping function.

The initial $k=0$ phonon-like chain set-up produces negligible initial amplitudes of the longitudinal waves. Longitudinal waves will only emerge in the presence of a resonant growth in the solution of Eq. \eqref{para_reso:hill-eq}. We expect the emerging wave to possess the mode number that yields 
the largest growth rate. 

Let us therefore analyse the stability of the solutions of Eq. \eqref{para_reso:hill-eq} and identify this value of $m$ depending on the system parameters. 
Through a low energy approximation the Hill equation \eqref{para_reso:hill-eq} can be transformed into a Mathieu equation whose stability can be analysed in an analytical manner, see \cite{MT_gross}. Here, we pursue an exact investigation by means of Floquet stability analysis. We can write Eq. (\ref{para_reso:hill-eq}) in the form 
\[\frac{d}{dt}\begin{pmatrix}\dot{\tau}\\ \tau \end{pmatrix}=\begin{pmatrix}0 &-c^2\left(1-\ddot{\widetilde{r}}\,^0(t)/ \left( c^2\,\widetilde{r}\,^0(t)\right) \right) \\ 1& 0  \end{pmatrix}\begin{pmatrix}\dot{\tau}\\ \tau \end{pmatrix}.\]
The pumping function is continuous and periodic (with period $\mathcal{T}=2\,\pi/\omega_0$). Thus, Floquet theory applies and postulates the existence of a transition matrix $\underline{\underline{\rho}}$ that projects the current state vector $\lbrace \tau(t), \dot{\tau}(t)\rbrace $ to the one after a period's time $\lbrace \tau(t+\mathcal{T}), \dot{\tau}(t+\mathcal{T})\rbrace $ . Therefore, the eigenvalues, $\mu_{k=\lbrace 1,2\rbrace}$, of $\underline{\underline{\rho}}$ determine the stability of the solution. It diverges if $|\mu_k|>1$ for at least one $k$. Then, the growth rate of the longitudinal wave pattern reads
\begin{align*}
\Gamma_{\varphi}=\frac{\ln \left( \underset{ k }{\max}\,\left|\mu_k\right| \right)} {\mathcal{T}}. 
\end{align*} 
To determine the transition matrix we (numerically) integrate Eq. (\ref{para_reso:hill-eq}) over one period for two linearly independent initial state vectors. Writing the initial state vectors as the columns of matrix $\underline{\underline{U_0}}$ and the state vectors after integration as columns of $\underline{\underline{U_1}}$ the transition matrix can be calculated from $\underline{\underline{U_1}}=\underline{\underline{\rho}}\,\underline{\underline{U_0}}\quad\Rightarrow\quad \underline{\underline{\rho}}=\underline{\underline{U_1}}\,\underline{\underline{U_0}}^{-1}.
$

In order to determine the longitudinal mode number we first choose a certain energy. This defines the amplitude of the radial stimulus according to Eq. \eqref{eq:isolate_A}. Then, at each point in parameter space we scan through the range of possible values of $m$ and determine the largest eigenvalue of the associated transition matrices. This value determines the growth rate of the longitudinal mode and the according $m$ constitutes its wave number, see Figs. \ref{fig:longi_Gamma} and \ref{fig:longi_m}. Larger energies lead to an increase of mode numbers and their growth rates and shrink the parameter regions where no longitudinal wave modes are expected. The identification of these regions is a particularly important feature of our theory because the absence of longitudinal waves has a crucial impact on the distribution of energy among the different degrees of freedom (see Fig.\,\ref{fig:energy_transf}) and thus on the escape behaviour.

\subsection*{Interplay of transversal and longitudinal wave modes}

\begin{table}
  \centering
\begin{tabular}{l c c c}
\hline

& $\lambda$ & $\kappa\,\Delta\Theta^2$ & description \\
\hline
 \PCa & $0.85$ & $1.25\cdot 10^{-4}$ & dominant radial mode, $m_r\approx 18$\\
\PCb & $0.4$ & $1.00$ & dominant longit. mode, $m_{\varphi}=2$\\
\PCc & $0.2$ & $1.10$ &dominant longit. mode, $m_{\varphi}=3$\\
\PCd & $0.4$ & $5.00\cdot 10^{-3}$ & disordered regime\\
\hline
\end{tabular}
  \caption{Parameter choices}
  \label{tab:PCs}
\end{table}

\begin{figure}
\centering
\textbf{Transversal Wave Modes}\\ \vspace{0.5\baselineskip}
\begin{subfigure}[t]{0.49\linewidth}
%\centering
%\includegraphics[height=0.118\paperheight]{./images/MI_m_max}
\includegraphics[height=0.12\paperheight]{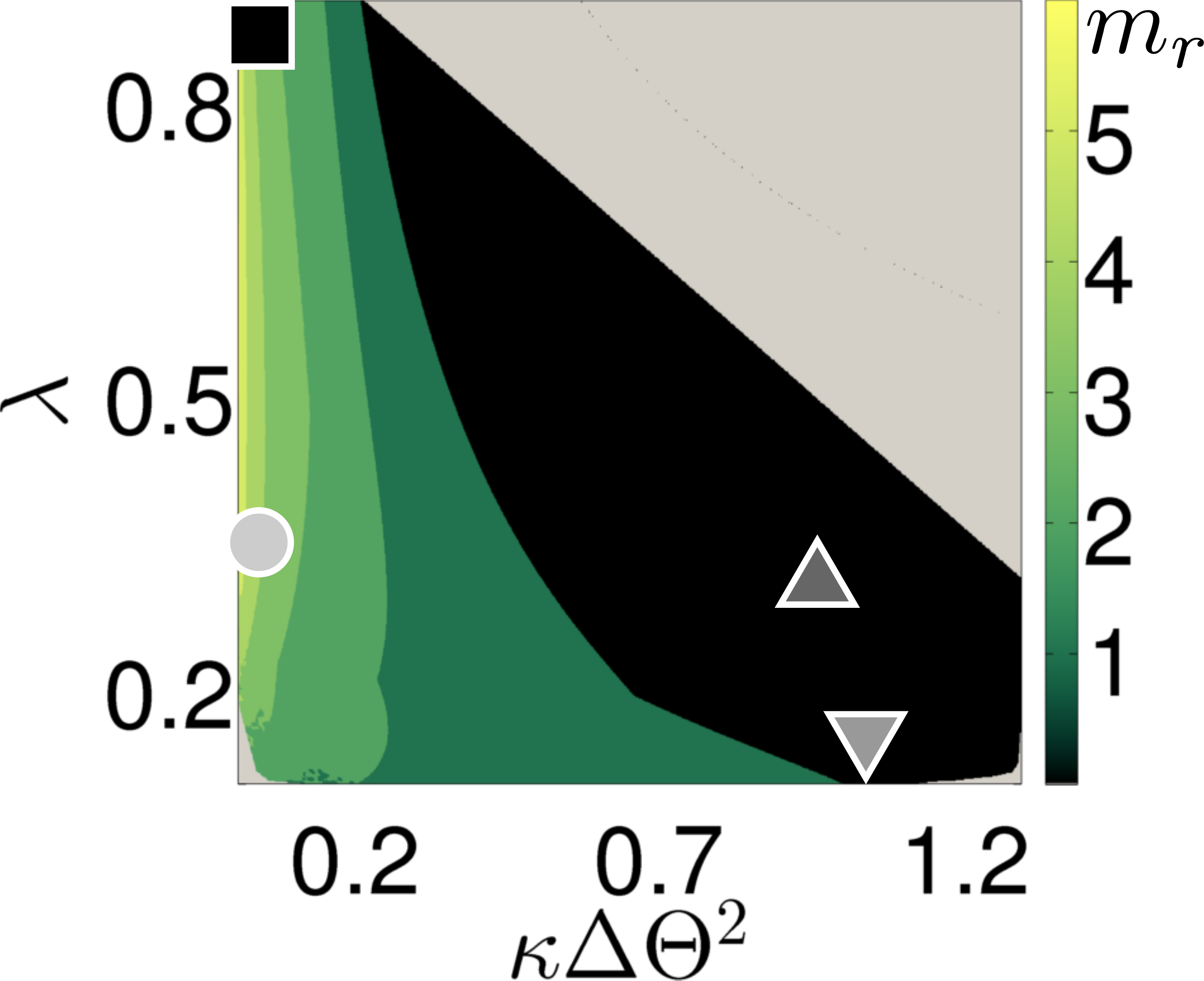}
\caption{Predicted mode number}
\label{fig:MI_m_max}
\end{subfigure}
\hfill
\begin{subfigure}[t]{0.49\linewidth}%4/7*0.48=0.56  3/7*0.98=0.42
\centering
\includegraphics[height=0.12\paperheight]{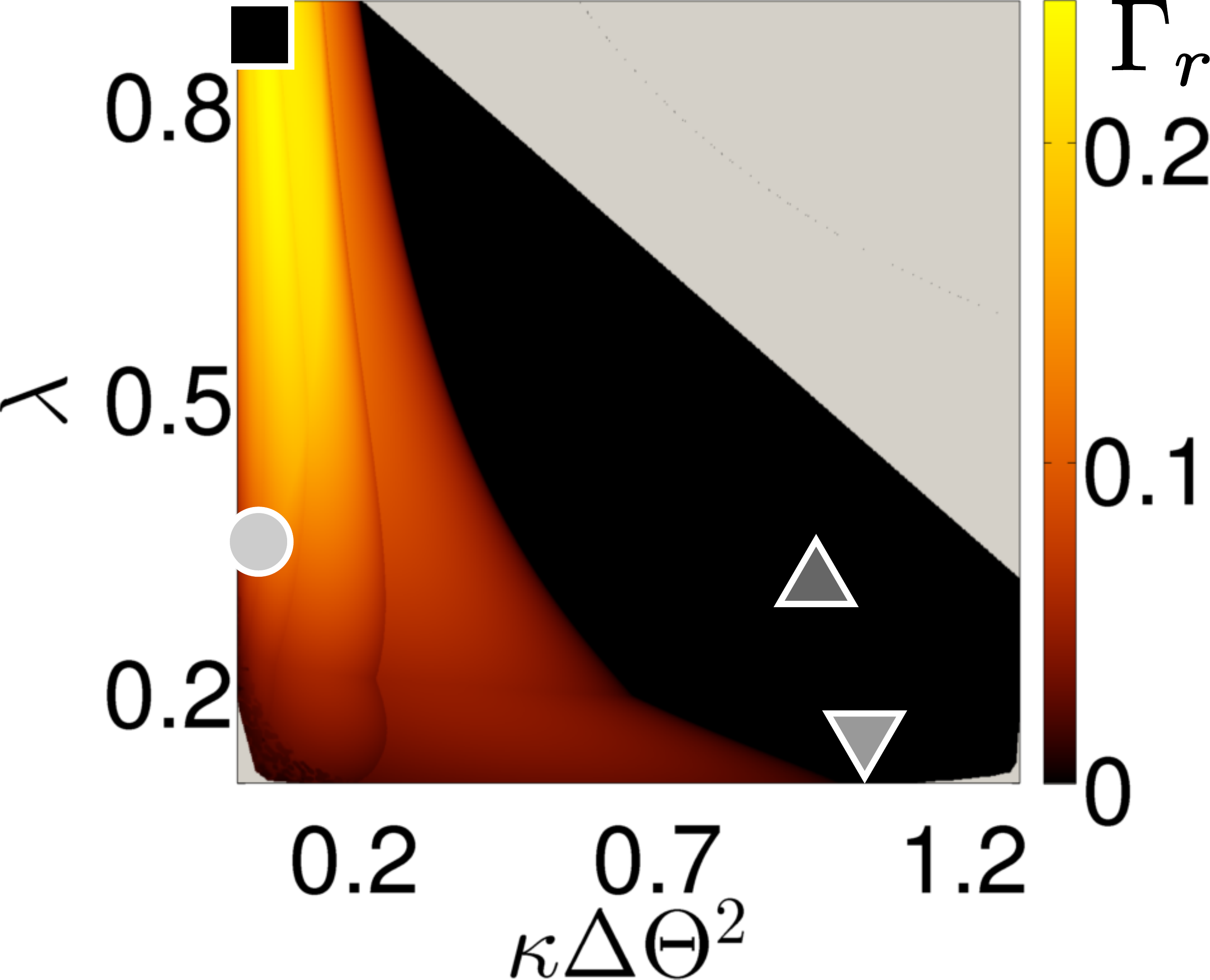}
\caption{Predicted growth rate}
\label{fig:MI_Gamma_max}
\end{subfigure}
\textbf{Longitudinal Wave Modes}\\ \vspace{0.5\baselineskip}
\begin{subfigure}[t]{0.49\linewidth}
%\centering
\includegraphics[height=0.12\paperheight]{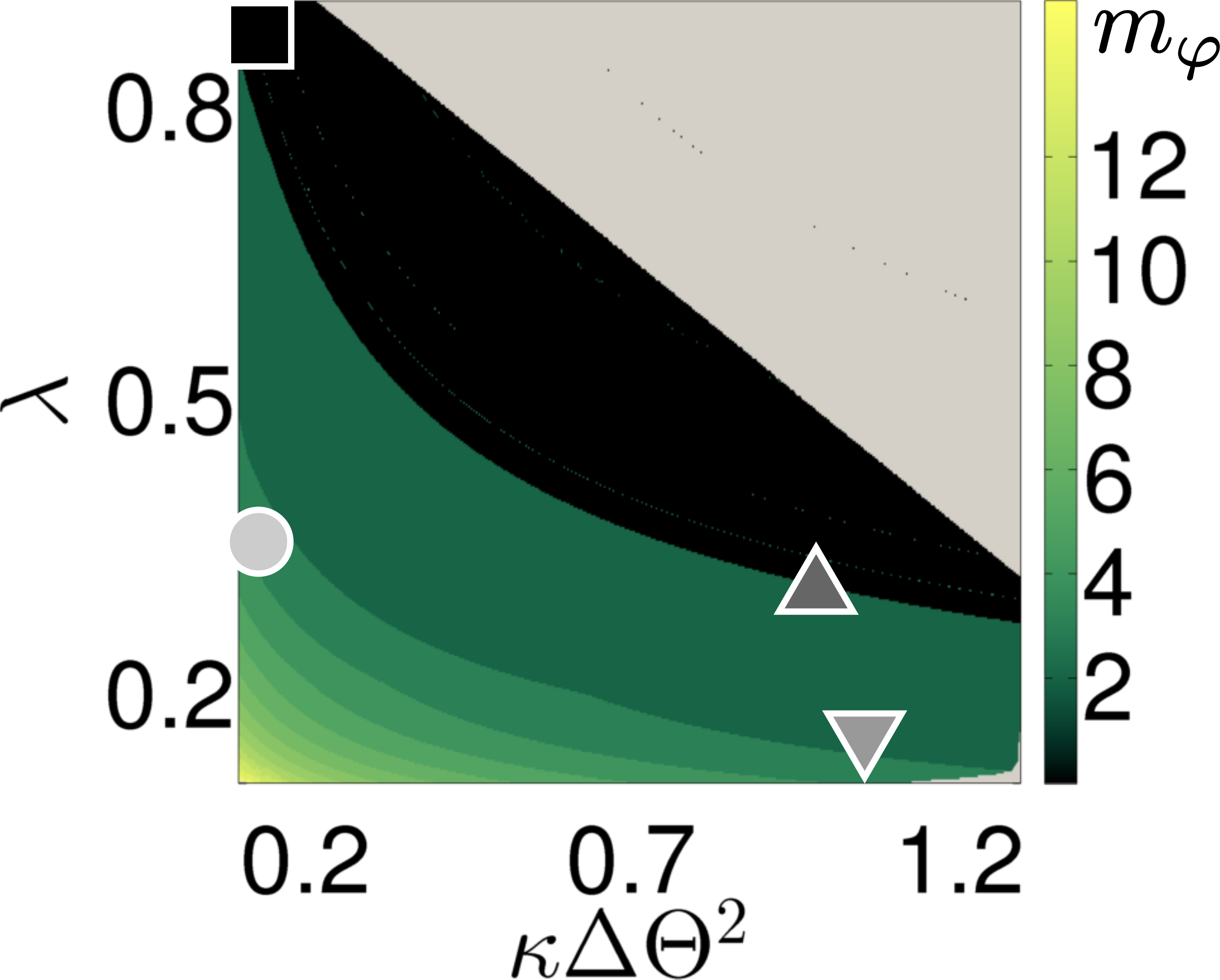}
\caption{Predicted mode number}
\label{fig:longi_m}
\end{subfigure}
\hfill
\begin{subfigure}[t]{0.49\linewidth}%4/7*0.48=0.56  3/7*0.98=0.42
\centering
\includegraphics[height=0.12\paperheight]{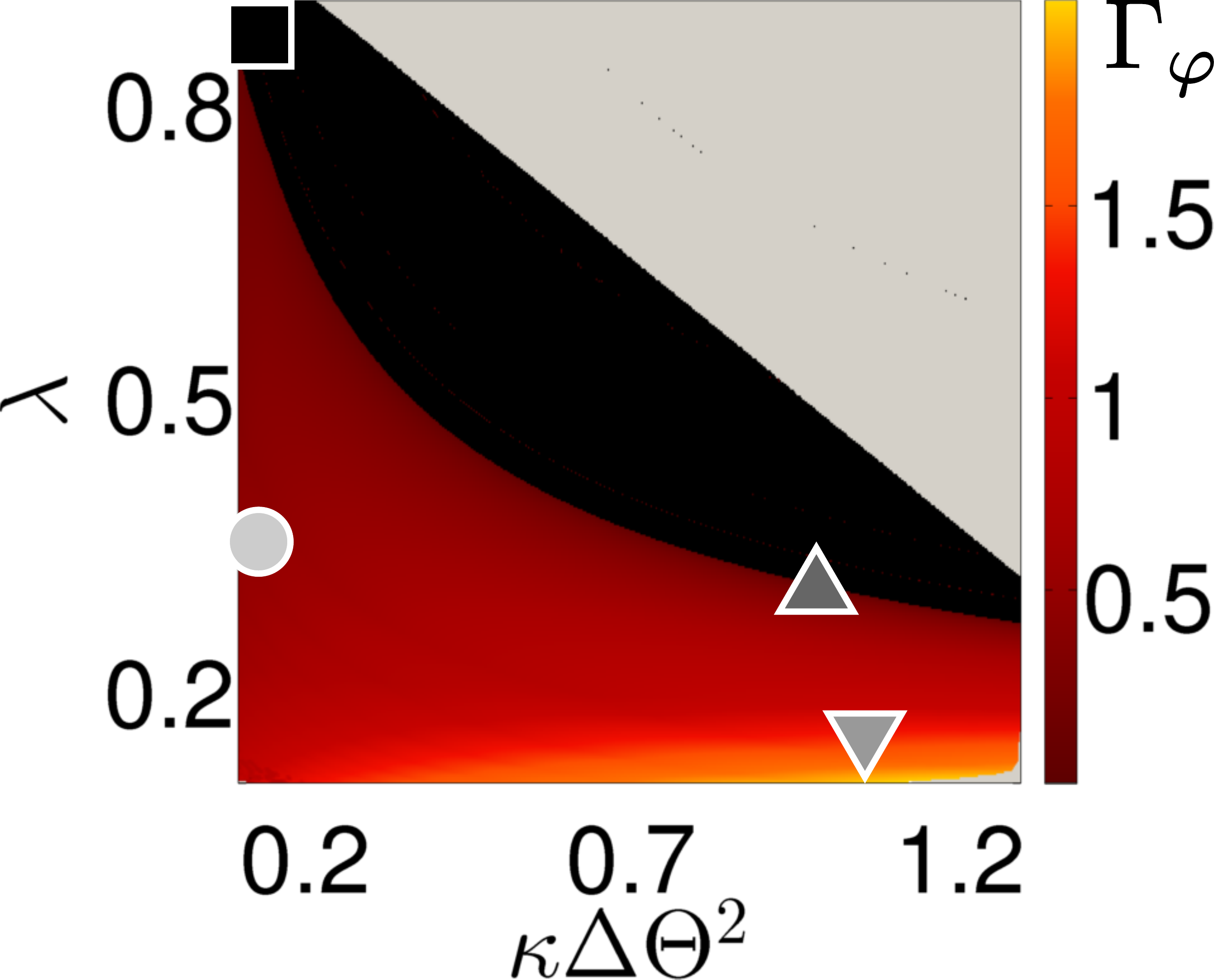}
\caption{Predicted growth rate}
\label{fig:longi_Gamma}
\end{subfigure}
\caption{Analytical results from the study of emerging wave modes. Symbols (\protect\PCa \protect\PCb \protect\PCc \protect\PCd) indicate parameter choices according to table \ref{tab:PCs}. $\epsilon_{\mathrm{scale}}=4$. No wave modes are expected in black areas, grey parameter areas are either excluded (Fig.\,\ref{fig:min_E_config_stability}) or their activation energy could not be determined (Fig.\,\ref{fig:E_act}). }
\label{fig:M_Gamma_results}
\end{figure}

The previous sections provide a theoretical framework describing the emergence of distinct wave patterns arising shortly after the initialisation of the system. The predicted wave mode numbers and rates are represented in Fig.\,\ref{fig:M_Gamma_results}. 

The energy values chosen for Fig.\,\ref{fig:M_Gamma_results} are a function of the parameters 
and their scaling properties in relation to the activation energy will be defined in the next section.

The two derivations, one for the transversal and one for the longitudinal modes, essentially relied on the elimination of a degree of freedom for each oscillator and additional approximations. Obviously, it is valid to ask how well those theoretical results correspond to the observations of the full system. Furthermore, the two different wave modes have been treated independently of each other so that is yet unclear whether they can appear simultaneously and if so, how they interact. 

To address these question we follow the temporal evolution of the full system by means of a numerical integration of its equations of motion. A high numerical accuracy is necessary to prevent significant energy deviations at different integration times. Therefore the equations of motion \eqref{qdyn1} have been integrated using a Runge-Kutta scheme of fourth order, choosing a time step  small enough (typically of the order of $10^{-4}$) to ensure that the energy deviation remains smaller 
than $10^{-12}$ throughout the entire simulation time.

\begin{figure}
\centering
\includegraphics[width=1\linewidth]{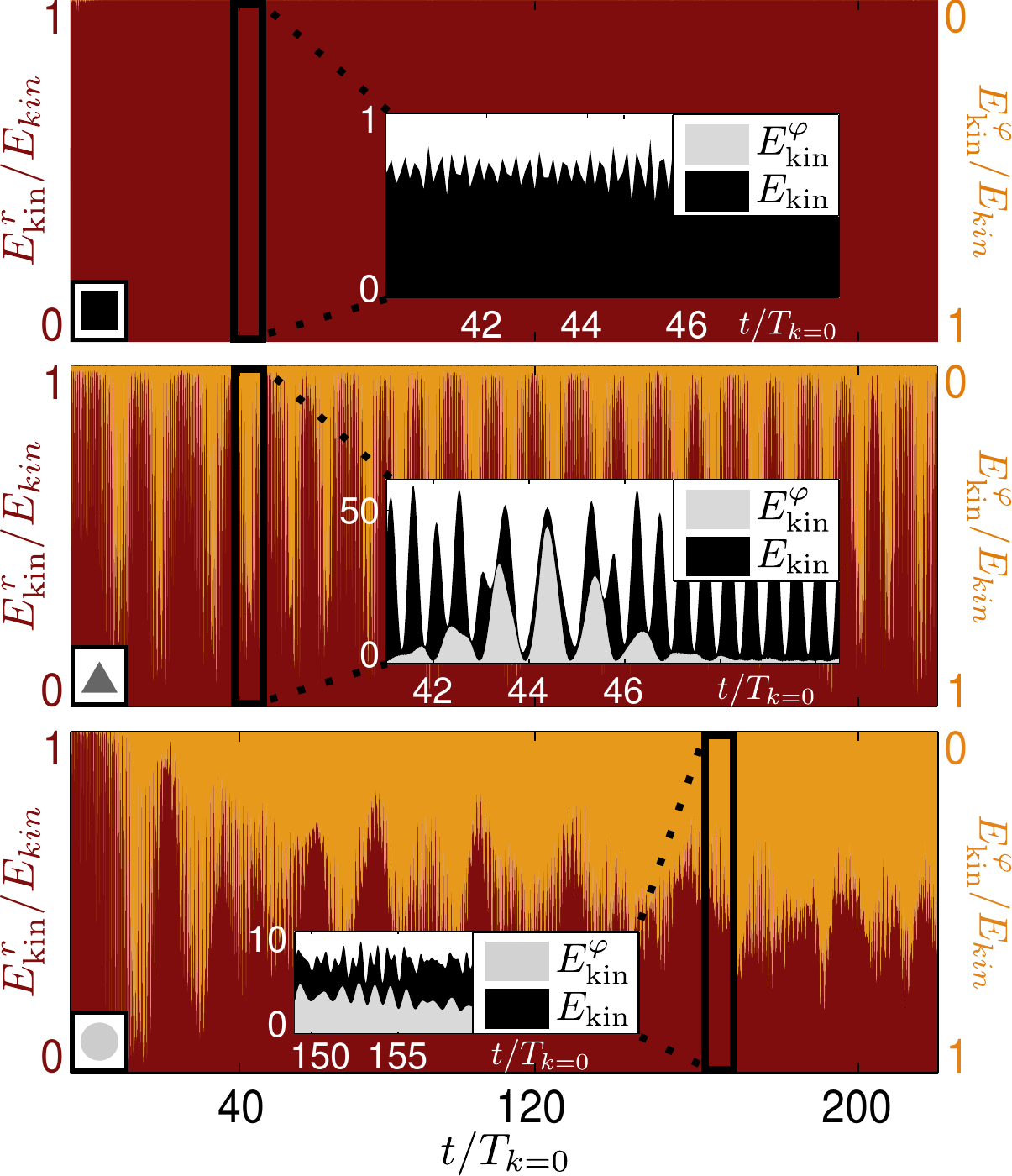}
\caption{The background plots depict how the total kinetic energy divides among the longitudinal (yellow) and radial (red) contribution, $E_{\mathrm{kin}}^{\,r}$ and $E_{\mathrm{kin}}^{\,\varphi}$, as a function of time, see Eq. \eqref{E_kin}. The insets show the temporal evolution of absolute kinetic energy values within the marked time frames. The three panels represent different parameter choices (indicated by the symbols \protect\PCa \protect\PCb \protect\PCd \  according to table \ref{tab:PCs}) which represent the three different dynamical regimes as described in the text. Time is measured in units of $k=0$ phonon periods $T_{k=0}=2\pi/\omega_0$. For all choices of parameters, $\epsilon_{\textrm{scale}}=2$.}
\label{fig:energy_transf}
\end{figure}

For different choices of parameter values we can identify three dynamical regimes  depending on which wave modes will emerge. This has a direct influence on how energy is transferred into different degrees of freedom. We can characterise the different dynamical states by tracking the radial and longitudinal contributions to the kinetic energy as illustrated in Fig.\,\ref{fig:energy_transf}
\begin{align}
\begin{split}
\label{E_kin}
E_{\mathrm{kin}}&=\sum_{i=0}^{N-1} \frac{\mathbf{p}_i^2}{2}=\sum_{i=0}^{N-1} \left( \left( \mathbf{p}_i\,\mathbf{e}_r \right)\mathbf{e}_r + \left( \mathbf{p}_i\,\mathbf{e}_{\varphi} \right)\mathbf{e}_{\varphi} \right)^2\\
&=\underbrace{\sum_{i=0}^{N-1} \left( \mathbf{p}_i\,\mathbf{e}_r \right)^2}_{\displaystyle E_{\mathrm{kin}}^{\,r}} +\underbrace{\sum_{i=0}^{N-1} \left( \mathbf{p}_i\,\mathbf{e}_{\varphi} \right)^2}_{\displaystyle E_{\mathrm{kin}}^{\,_{\varphi}}}. 
\end{split}
\end{align}

The system's behaviour can be dominated by transversal wave modes. This is the case when longitudinal wave modes are absent (black areas in Figs.\,\ref{fig:longi_m} and \ref{fig:longi_Gamma}) or when their mode number is high ($m_{\varphi}>20$) because then the angular displacement of individual oscillators (wave amplitude) becomes small so that the assumption of fixed angular components in the derivation of the modulation instability becomes increasingly justified. Such a setting will (even for large integration times in the order of the duration of many hundred phonon oscillations) prevent an energy transfer into longitudinal degrees of freedom (see ''\PCa '' in Fig.\,\ref{fig:energy_transf}: Virtually all kinetic energy is contained in the radial motion, due to the emergence of numerous, unsynchronised breathers the total kinetic energy remains approximately constant).

In the converse case where radial wave modes are absent or of high wave number ($m_r>20$) the system's behaviour will be dominated by longitudinal wave modes. As it became clear in the previous section, these modes arise due to a resonant excitation from the initial phonon mode. With the growth of the longitudinal mode the system energy is transferred from the radial into the longitudinal degrees of freedom. However once the phonon oscillations are attenuated the longitudinal modes are no longer excited so that in turn their amplitude will decrease, thereby leading the system back into the phonon-like state. This again triggers the re-emergence of the longitudinal wave. The system therefore resides in an oscillatory regime (see ''\PCb'' in Fig.\,\ref{fig:energy_transf}: In periodically repeating intervals nearly all of the kinetic energy is transferred into the longitudinal motion. Both the phonon and longitudinal mode cause a synchronous oscillation between kinetic and potential energy for each oscillator. This causes the oscillations of the total kinetic energy).

In the simultaneous presence of radial and longitudinal wave modes the system will evolve into a highly disordered state. The mode of larger growth rate will emerge first but will soon start mixing with the other mode such that the system develops a nonspecific, irregular long-term behaviour. Once the initial wave patterns have died out, the resulting disorder leads to a more homogeneous distribution of the system energy into all degrees of freedom (see ''\PCd '' in Fig.\,\ref{fig:energy_transf}: After some initial longitudinal oscillations the energy distribution becomes effectively unstructured and gets close to equipartition).

Parameter choices that represent the three different regimes can be found in table \ref{tab:PCs}.

\section{Deterministic escape\label{sect:det-esc}}

So far this analysis has focused on the description of emerging wave modes. However, we recall that the chain is initially placed in a meta-stable state. This brings about yet another level of investigation which concerns the escape of the chain beyond the potential barrier into the unbounded regime. The randomly perturbed initial conditions along with the chaotic dynamics supports a treatment of this first-passage time problem in statistical terms. 
Herein, we will focus on the role that the emerging wave modes play for the escape behaviour.

\subsection*{Transition states and escape channels\label{sect:transition_state}}

\begin{figure}
\centering
\begin{subfigure}[t]{0.32\linewidth}
\includegraphics[height=0.125\paperheight]{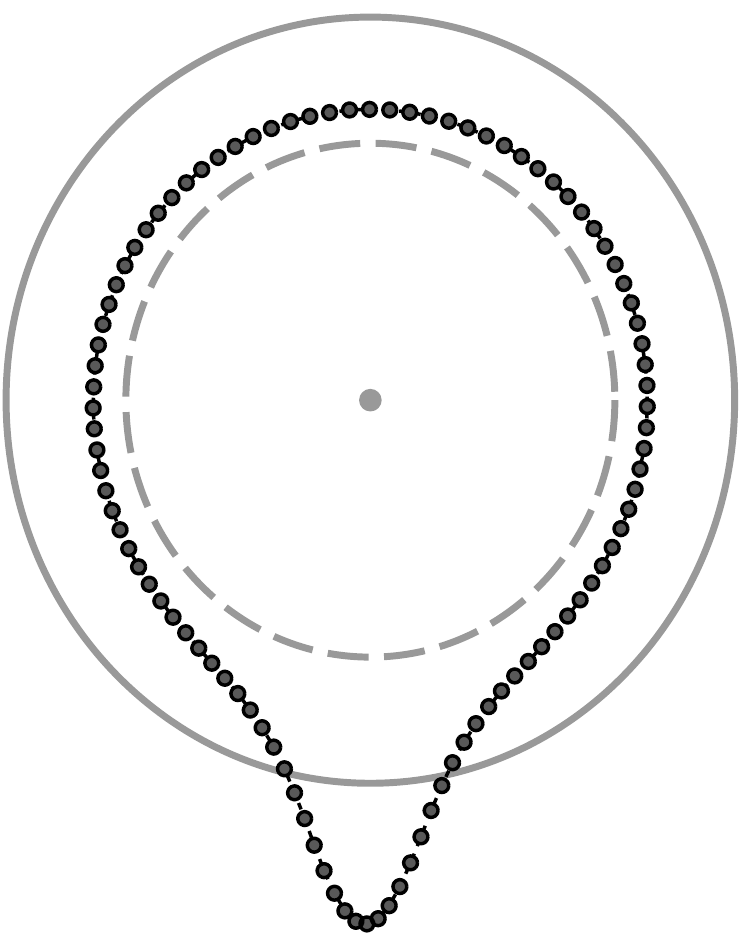}
\caption*{type I}
\end{subfigure}
\hfill
\begin{subfigure}[t]{0.32\linewidth}
\includegraphics[height=0.125\paperheight]{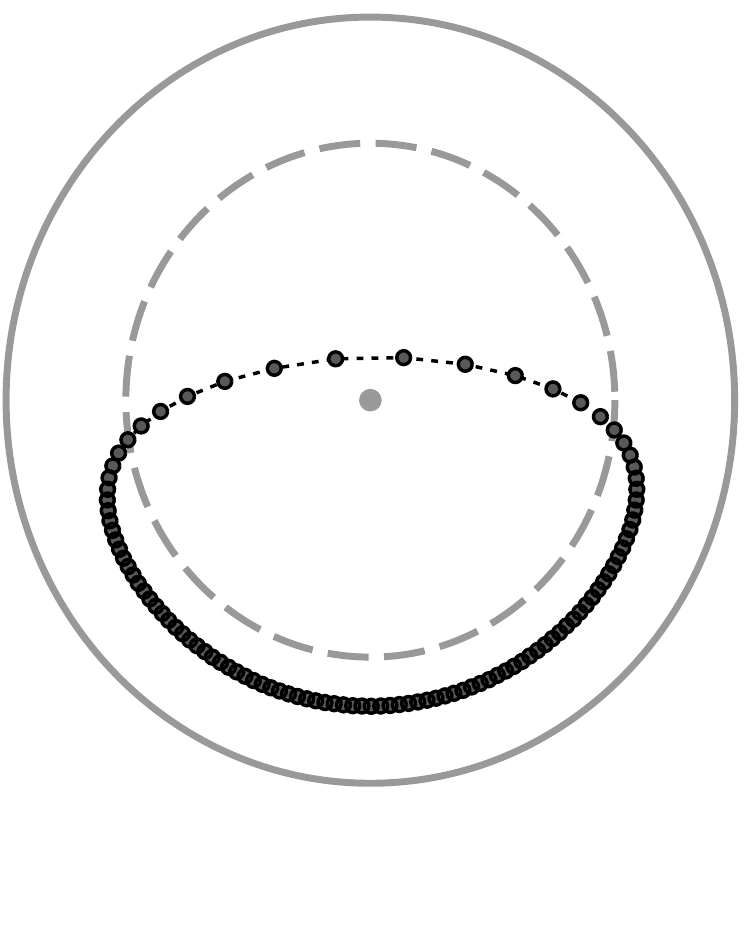}
\caption*{type IIa}
\end{subfigure}
\hfill
\begin{subfigure}[t]{0.32\linewidth}
\includegraphics[height=0.125\paperheight]{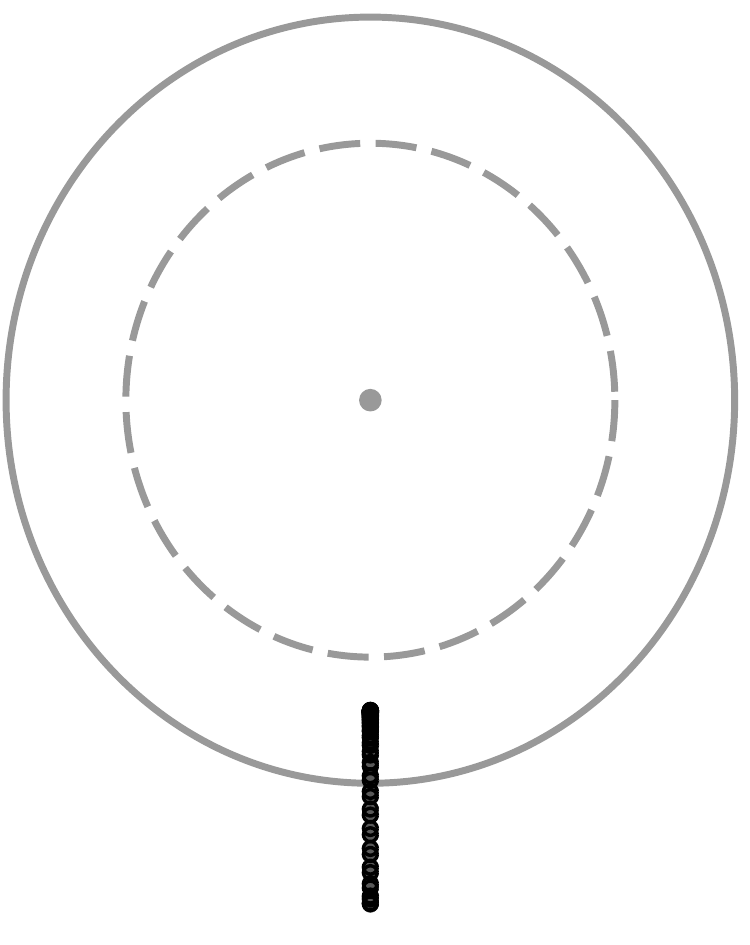}
\caption*{type IIb}
\end{subfigure}
\caption{Relevant transition state types.}
\label{fig:overview_ts}
\end{figure}

It is essential to know the minimal energy necessary for the system to leave the potential energy basin for another region in configuration space, in other words the minimum energy needed for an escape event. This question leads to the search for first-order saddle points of the potential energy surface. From the definition of the Hamiltonian, Eq. \eqref{scaledHam} follows the expression of the potential energy
\[
U\left(\left\lbrace\mathbf{q}_i \right\rbrace \right)= \sum_{i=0}^{N-1}\left[\frac{\kappa}{2} \left(\mathbf{q}_i-\mathbf{q}_{i+1}\right)^2-\sqrt{\mathbf{q}_i^2}+\cos\left( \frac{\sqrt{\mathbf{q}_i^2}}{\lambda}\right)\right].
\]
Thus, we solve
$
\nabla U\left(\left\lbrace\mathbf{q}_i \right\rbrace \right)=0,
$
such that the Hessian matrix of U
%\begin{align*}
%H_{o,p}\left(\left\lbrace\mathbf{q}_i \right\rbrace \right)=\frac{\partial}{\partial q_o}\frac{\partial}{\partial q_p}U\left(\left\lbrace\mathbf{q}_i \right\rbrace \right) \quad o,p=\lbrace x,y \rbrace , \lbrace i=0,1\ldots (N-1) \rbrace ,
%\end{align*} 
has only positive eigenvalues except for a single negative one. An algorithm originating from theoretical chemistry, the dimer method \cite{TS:henkelman,TS:olsen,TS:heyden,TS:pedersen}, proved effective to solve this numerically difficult task. It identifies a manifold of different saddle points of which most have no relevance to our study because they are either high-energy configurations, unattainable by our setting, or configurations that have already escaped from the meta-stable initial setting. Examples of the three remaining so called transition state types are shown in Fig.\,\ref{fig:overview_ts}. 

\begin{figure}
\centering
\parbox{0.32\linewidth}{time: 90.00}\hfill
\parbox{0.32\linewidth}{time: 99.45}\hfill
\parbox{0.32\linewidth}{time: 99.90}\\ 
\vspace{1ex}
\includegraphics[width=0.32\linewidth]{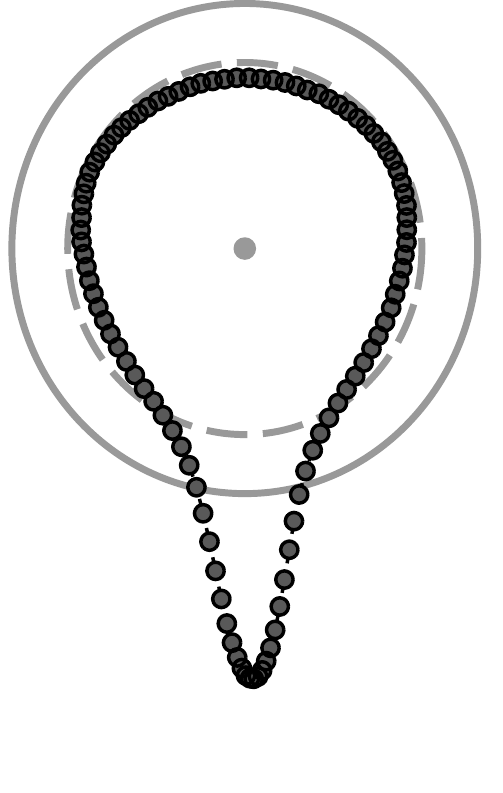}\hfill
\includegraphics[width=0.32\linewidth]{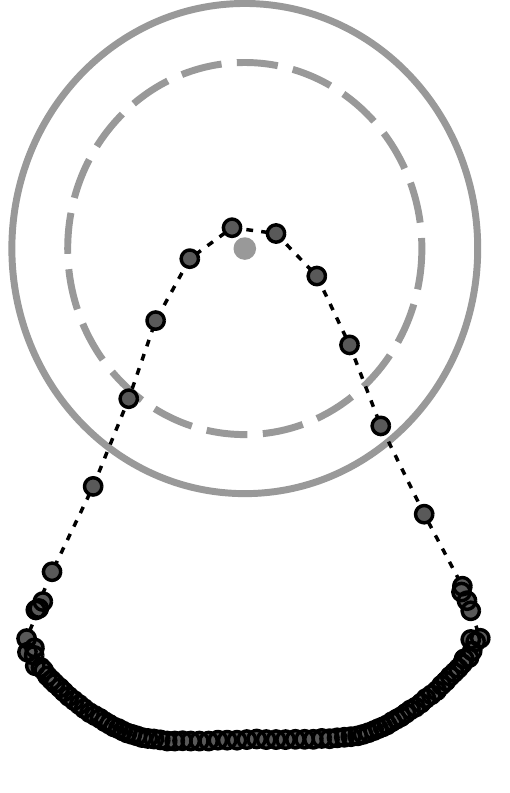}\hfill
\includegraphics[width=0.32\linewidth]{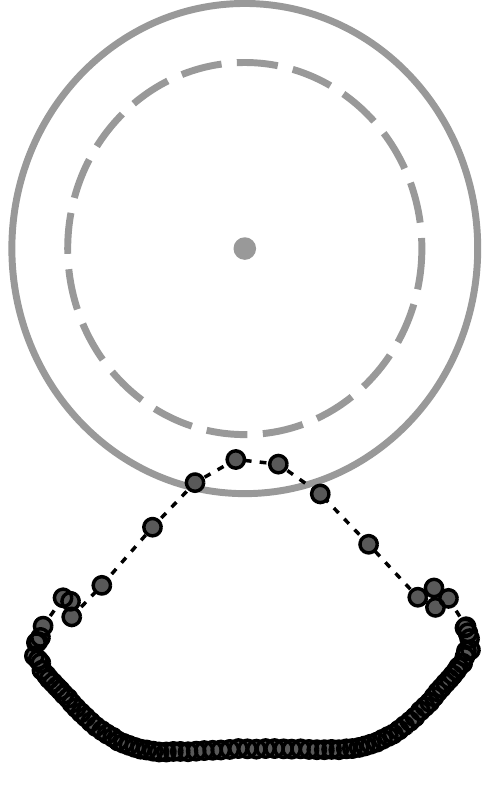}
\caption{Snapshots of an escape process of type I. Parameters: $\lambda=0.8$, $\kappa\,\Delta\Theta ^2=0.08$%, , $E=20$.
}
\label{fig:I_escape}
\end{figure}

\begin{figure}
\centering
\parbox{0.32\linewidth}{time: 29.55}\hfill
\parbox{0.32\linewidth}{time: 33.15}\hfill
\parbox{0.32\linewidth}{time: 33.80}\\ 
\vspace{1ex}
\includegraphics[width=0.32\linewidth]{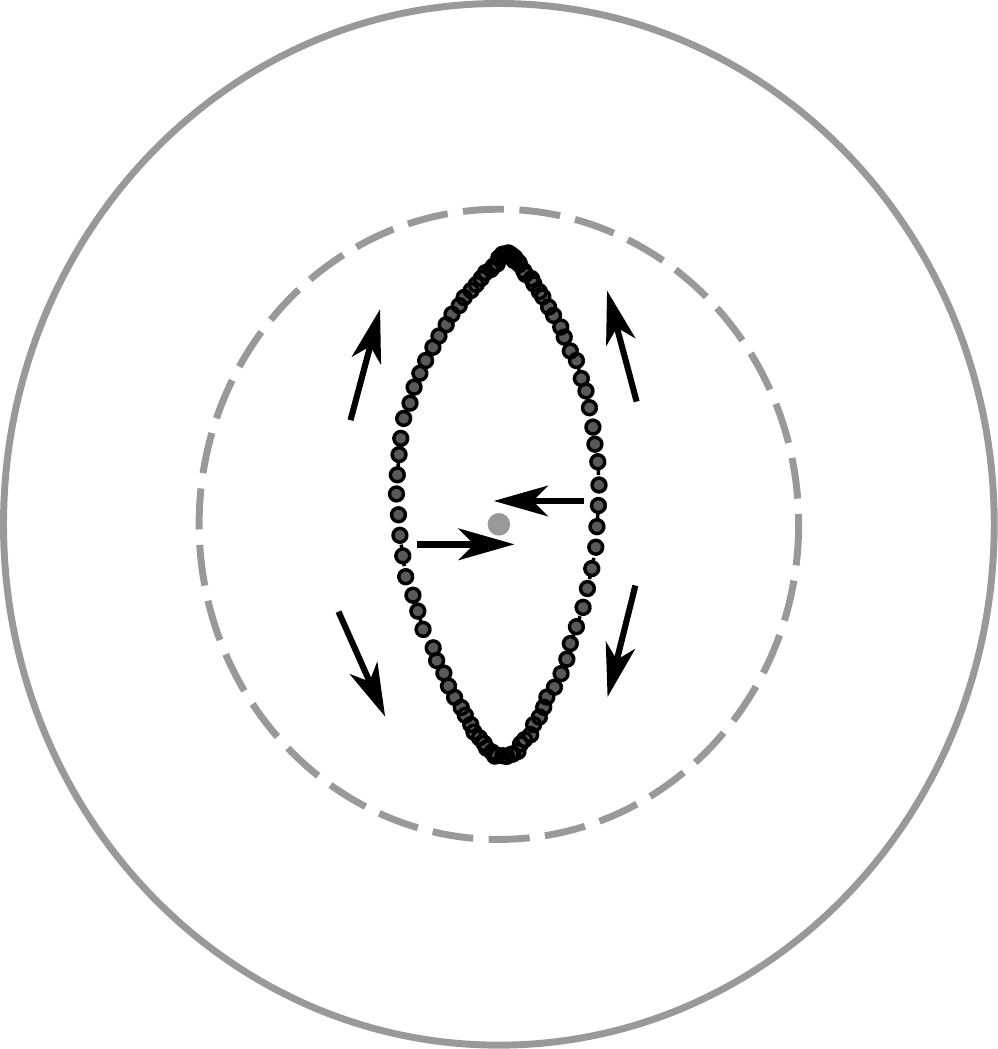}\hfill
\includegraphics[width=0.32\linewidth]{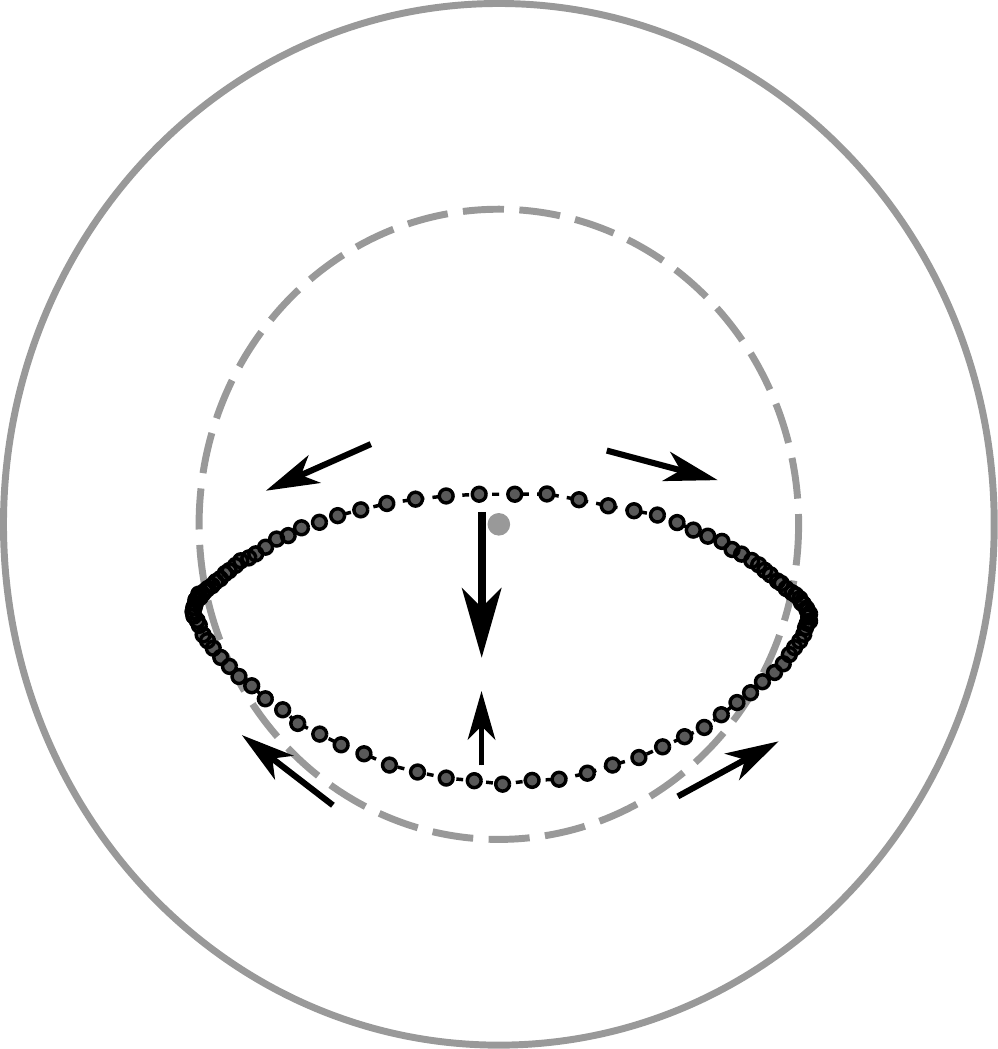}\hfill
\includegraphics[width=0.32\linewidth]{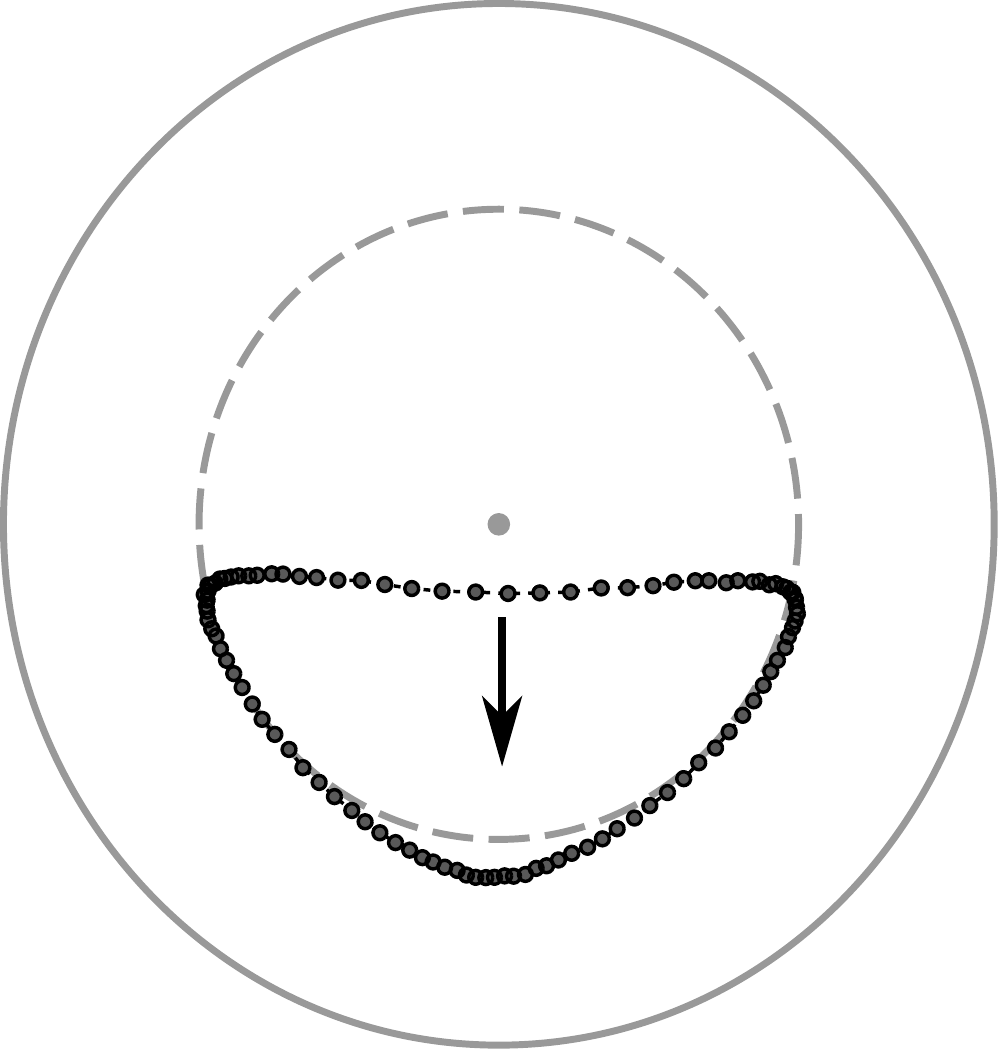}\\
\vspace{2ex}
\parbox{0.32\linewidth}{time: 35.20}\hfill
\parbox{0.32\linewidth}{time: 35.60}\hfill
\parbox{0.32\linewidth}{time: 37.45}\\ 
\vspace{1ex}
\includegraphics[width=0.32\linewidth]{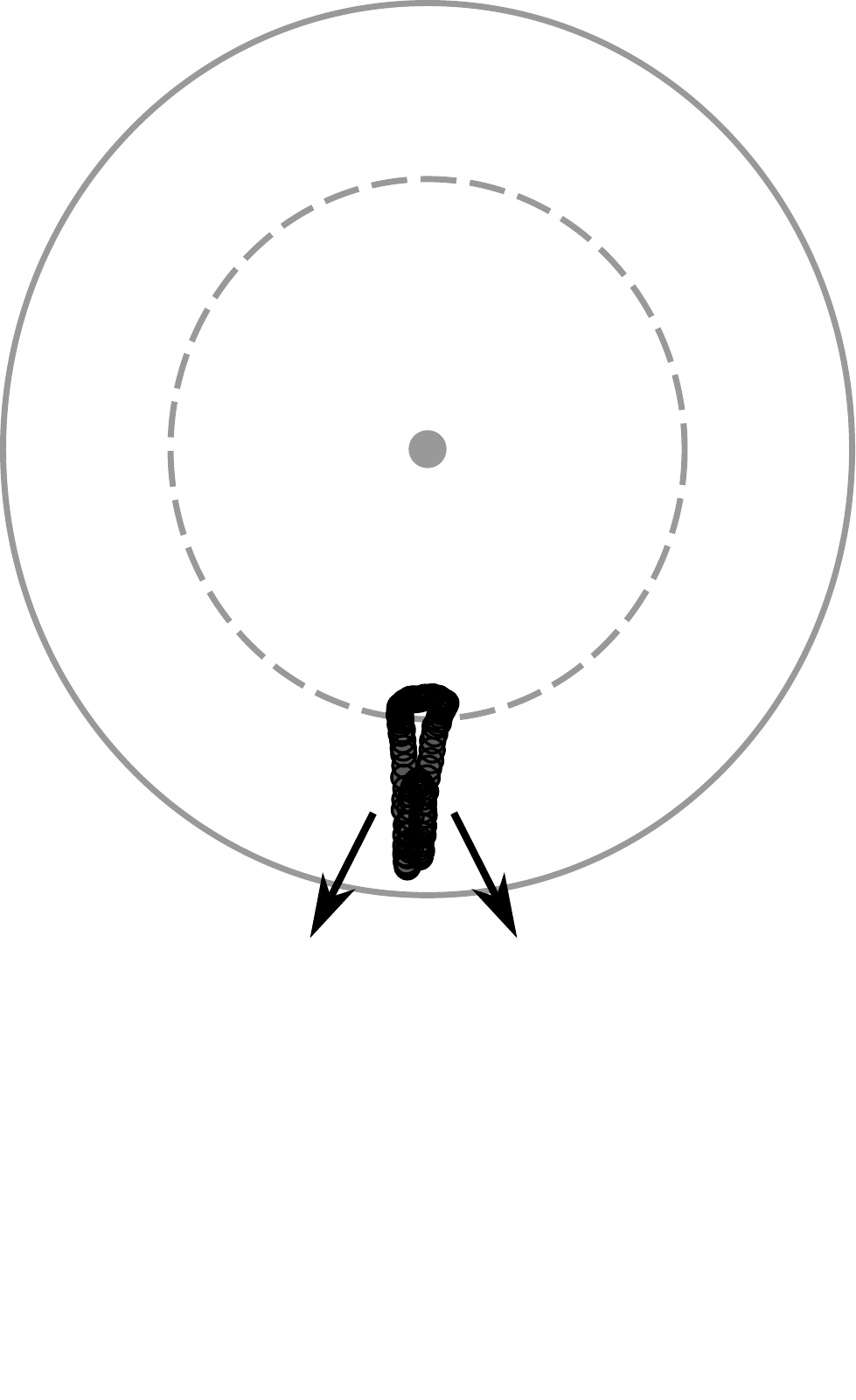}\hfill
\includegraphics[width=0.32\linewidth]{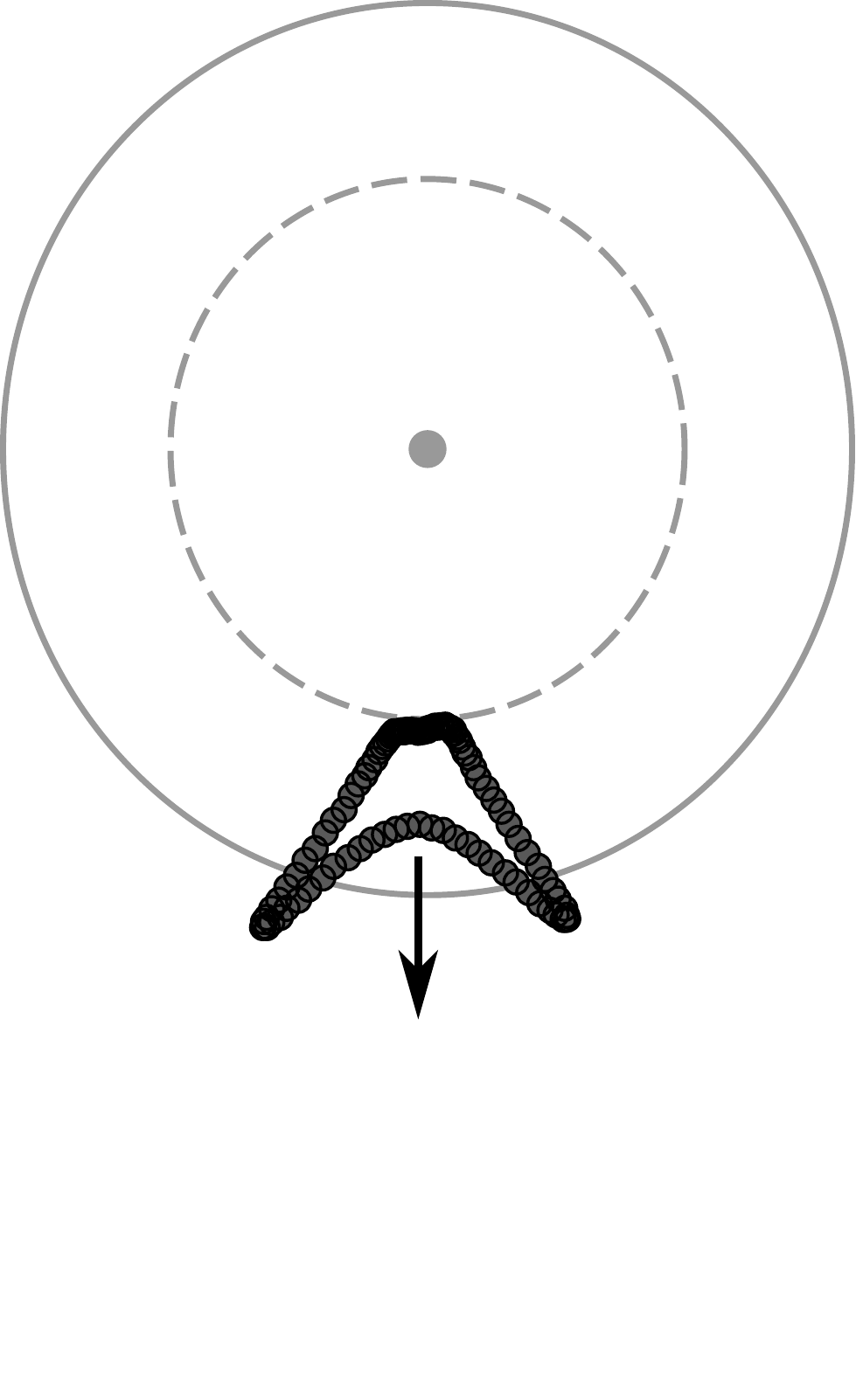}\hfill
\includegraphics[width=0.32\linewidth]{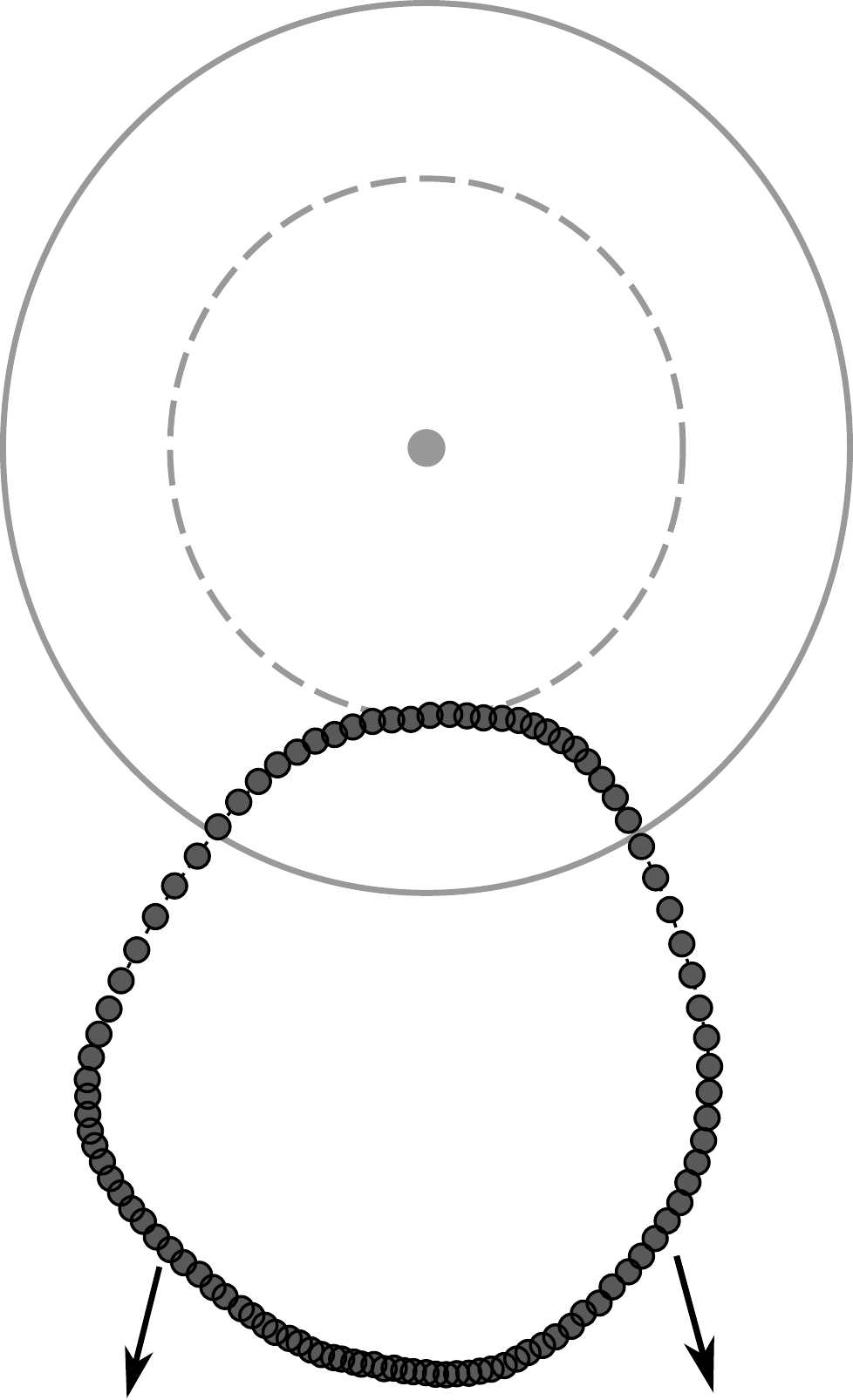}
\caption{Snapshots of an escape process of type II. Arrows indicate the chain movement. Parameters: ''\protect\PCb '' according to table \ref{tab:PCs}.}
\label{fig:II_escape}
\end{figure}

The existence of these transition states entails two different types of escape mechanisms. The escape related to transition state type I, as shown in Fig.\,\ref{fig:I_escape}, indicates a process in which a few oscillators surmount the potential barrier, are driven further down the outer slope of the potential barrier and thereby pull out the entire chain from the meta-stable state. An escape of type II, depicted in Fig.\,\ref{fig:II_escape}, describes the process in which the chain first surmounts the central potential hump, passing the transition state of type IIa, and then overcomes the potential barrier in the way indicated by transition state type IIb.

\begin{figure}
        \centering
    \includegraphics[height=.125\paperheight]{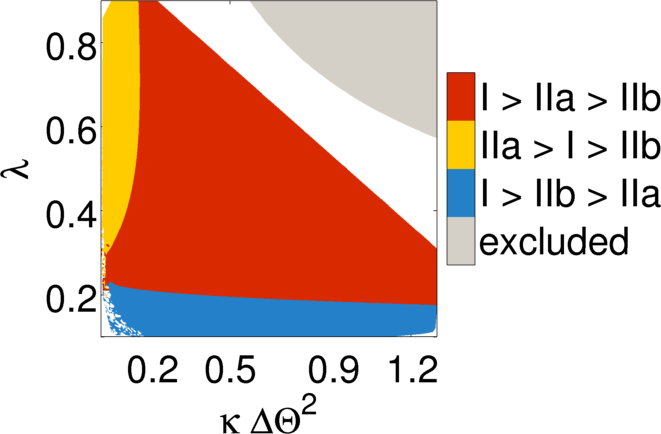}
                \caption{Comparison of the energies of different transition state types (white: no energy values determined, grey: excluded from parameter space).}
                \label{fig:comp_E}
\end{figure}
\begin{figure}
 \includegraphics[height=0.125\paperheight]{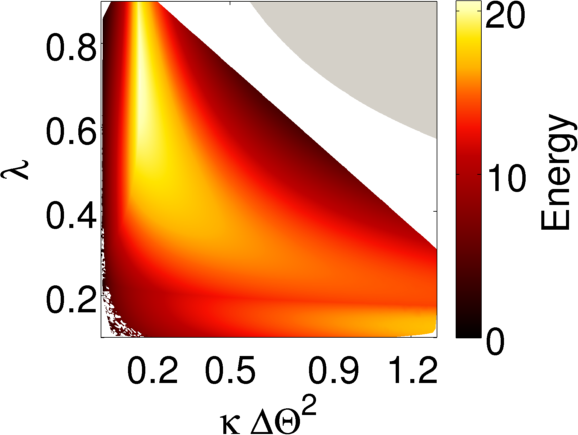}
  \caption{Activation energy, corresponds to $\epsilon_\mathrm{scale}=1$ (white: no energy values determined, grey: excluded from parameter space).}
                \label{fig:E_act}
\end{figure}

The minimal energy necessary for an escape through channel I is the energy content of the transition state of type I. For an escape through channel II it is the larger one of the transition states energies of type IIa and IIb because the phase space trajectory will have to pass through the vicinity of both transition states in order to leave the bounded regime. 

Let us thus compare the energy values of different transition states (Fig.\,\ref{fig:comp_E}). This enables us to determine the minimal escape energy (Fig.\,\ref{fig:E_act}), which will be referred to as the activation energy, $E_{\mathrm{Act}}$. It is computed as the energy difference between the associated transition state and that of the minimum energy configuration $E_0$. Thus,  $E_{\mathrm{Act}}=\mathcal{H}\left(\left\lbrace\mathbf{p}=0\right\rbrace,\left\lbrace\mathbf{q}=\mathbf{q}^*\right\rbrace\right)-E_0$, where $\lbrace\mathbf{q}^*\rbrace$ denotes the relevant transition state configuration corresponding to the escape channel of lower 
energy as described above. This notion allows to define the energy scaling parameter
\begin{align*}%\label{eq:eps_scale}
\epsilon_{\mathrm{scale}}=\frac{E-E_0}{E_{\mathrm{Act}}}.
\end{align*}

From now on, energies will only be expressed in terms of $\epsilon_\mathrm{scale}$, that is the ratio of the system's energy to the activation energy for a given set of parameters. This will not only allow to evaluate the significance of the energy value with respect to the escape process but also to meaningfully compare energy values for different choices of parameter values, which is cumbersome using absolute energy values because they could be anything from a small fraction to a multiple of the activation energy, depending on the parameter values.

However, we need to be aware that this definition of $\epsilon_\mathrm{scale}$ 
has consequences for the interpretation of the escape behaviour. There might be parameter values for which an escape through a certain channel will be very likely to occur as soon as the system's energy slightly exceeds this channel's transition energy. Such a setting could be interpreted as being favourable for the deterministic escape. However, if the transition energy of the other channel is 
significantly lower but this channel's escape mechanism is inhibited by the chain's dynamics, the chain will only escape for high values of $\epsilon_\mathrm{scale}$, so that such a choice of parameter values would still be seen as hampering a deterministic escape. 
%Of course this decision of interpreting energy values cannot be rigorously defended. It is a choice that seems most plausible and according to the findings in the next section the scenario mentioned above is atypical for most of the parameter space.

Note that without considerable additional efforts the dimer method does not converge throughout the entire parameter space. Points for which no energies could be determined are represented as white space in Fig.\,\ref{fig:comp_E} and Fig.\,\ref{fig:E_act}. However these regions are of no particular interest in the following considerations so that a closer examination is dispensable.

\subsection*{Escape times and characteristics}

Figure \ref{fig:comp_E} shows that for the largest portion of the relevant parameter space (red and blue areas) an escape through channel II is energetically more favourable. However, Sect. \ref{sec:waves} revealed how different parameter values lead the system into different dynamical regimes which in turn influence the escape behaviour. To study the interplay between transition state energies, wave modes and deterministic escape quantitatively, let us measure the escape times for the set of parameter choices given in table \ref{tab:PCs}, as they represent a broad spectrum of the possible dynamical regimes. 

\begin{figure}
        \centering
         \includegraphics[width=\linewidth]{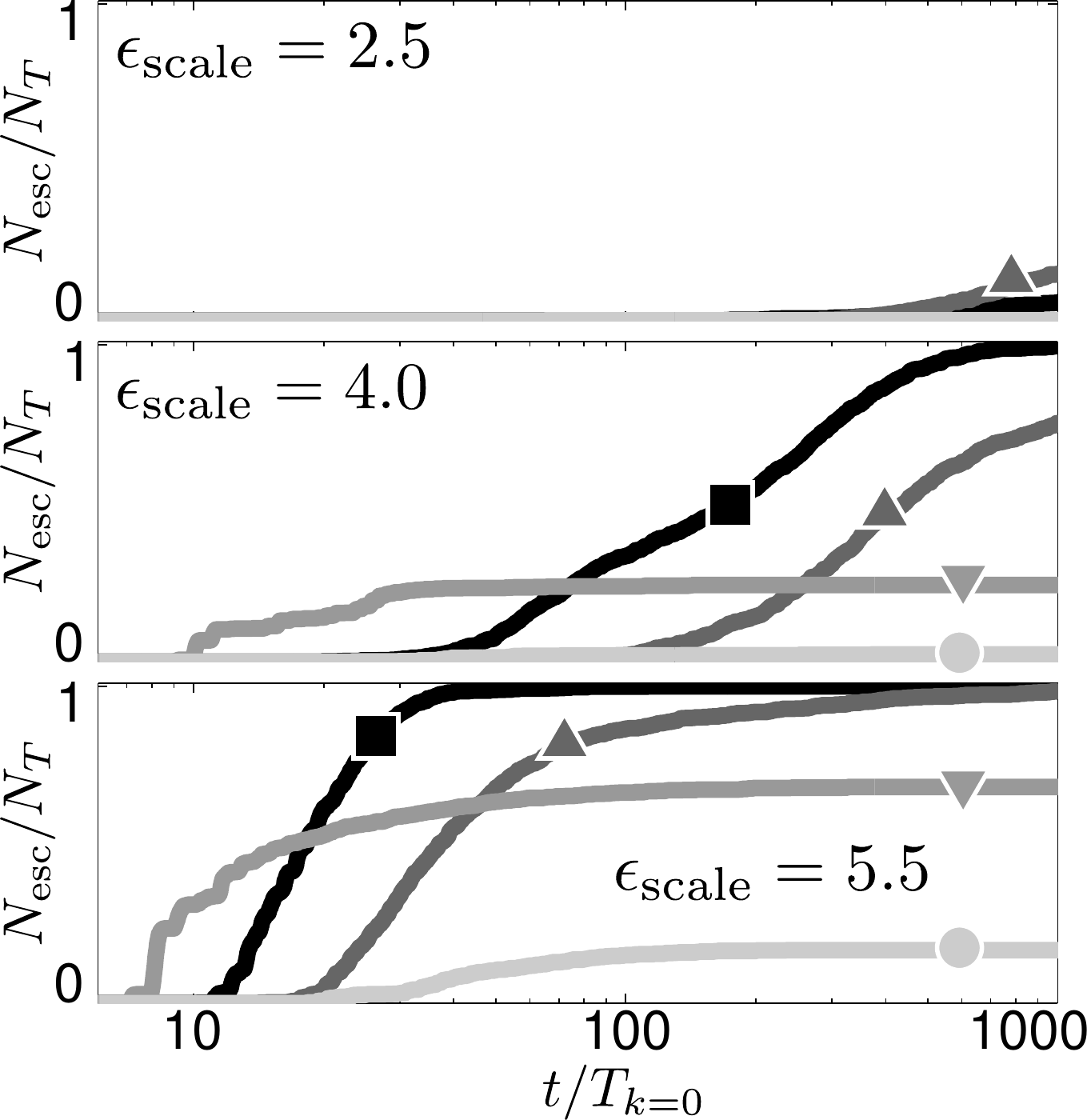}
         \caption{Cumulative escape time distributions. Parameters (\protect\PCa: $m_r\approx 18$, \protect\PCb: $m_{\varphi}=2$, \protect\PCc: $m_{\varphi}=3$, \protect\PCd: disordered regime) according to table \ref{tab:PCs}.}
         \label{fig:esc-distr}
\end{figure}

To this end, we prepare ensembles of systems with equal parameter values and energies. The random perturbations of their initial conditions create different chain realisations that produce a statistical ensemble of escape times proper for a meaningful interpretation of the escape process. Those are represented as cumulative escape time distributions in Fig.\,\ref{fig:esc-distr}, which depicts the fraction of chains 
that have escaped up until a certain time out of the total number of chain realisations, $N_T$. For each parameter choice and each value of $\epsilon_{\mathrm{scale}}$ we set $N_T=1000$. The escape time is defined as the time it takes from the system's initialisation until all oscillators have surpassed the potential barrier. To make escape times comparable between different sets of parameters we will measure them in units of $T_{k=0}=2\pi/\omega_0$, that is, the duration of $k=0$ - mode phonon-like oscillations.

It becomes clear from Fig.\,\ref{fig:esc-distr} that different dynamical regimes relate to a different escape behaviour. When the system's dynamics comprises a dominant transversal wave mode (\PCa), it forms breathers that promote an escape of type I, as shown in e.g. in Fig.\,\ref{fig:MI_high_wavenumber_escape}. Just as was observed in previous studies on deterministic escape of oscillator chain systems \cite{1d-chain_escape_EPL,1d-chain_escape_PhysRevE,FugHe07,fugman_etc_epj,martens08}, breathers have the tendency to concentrate energy in small chain segments. The creation of these highly excited segments can produce critical oscillator elongations in radial direction beyond the potential barrier that trigger an efficient escape of type I.

In the presence of dominant longitudinal modes (\PCb , \PCc) we observe an entirely different behaviour with significant distinctions depending on the wave number $m_{\varphi}$.  In general, an escape of type I is not expected because of the lack of energy concentration into critical radial elongations. However we see an enhancement of an escape of type II if $m_{\varphi}=2$ (\PCb). In such a case the chain is strongly stretched in between the two wave nodes. It tends to reduce the tension by decreasing the length of the stretched sections, which makes these more straight and they thereby surmount the central potential hump. The initial perturbations can break the symmetry of the longitudinal pattern which can cause one of the two stretched segments to overcome the potential hump. Exactly this can be observed in the first three snapshots of Fig.\,\ref{fig:II_escape}. The first two show the two wave nodes (first vertically then horizontally aligned) and the third  displays how the upper part of the stretched 
chain segment is carried over the potential hump. Generally, the subsequent trespass over the potential barrier is directly achieved due to the inert motion of the segment that was accelerated down the potential hump (this is because the energy of transition state type IIb is small whenever $m_{\varphi}=2$). All in all, this mechanism triggers an enhanced escape of type II. 

For dominant longitudinal modes of other wave numbers -- already for $m_{\varphi}=3$ (\PCc) -- we no longer find this special geometry of the longitudinal wave and even for large energies an escape is inhibited. We do see initial escapes in this case but those are only due to a high initial order. When the $m_{\varphi}=3$ mode is still mostly unperturbed it can cause some oscillator elongations in radial direction which can initiate an escape of type I. Later on, when the longitudinal mode is more irregular, those do not occur anymore so that the cumulative escape distribution (of \PCc) completely saturates in the long run.

Finally, a system that evolves towards a highly disordered state (\PCd), due to the mixing of longitudinal and transversal modes, also disperses the energy into all degrees of freedom (see bottom plot in Fig.\,\ref{fig:energy_transf}). Thus, there is neither an energy concentration into transversal degrees of freedom causing an escape of type I nor the special geometry of the longitudinal mode bringing about an escape of type II. Again, we find initial escape events that take place before the modes have mixed, their type depending on which mode has a larger growth rate (here mostly escape type I). But once the disordered state is achieved only rare, coincidental, critical elongations can drive the chain beyond the potential barrier. Hence, the escape is clearly inhibited compared to all other dynamical regimes.

Regarding the relation between the relevant escape channel and the different transition states energies, we find that parameter choices leading to dominant modes (for radial modes those are mostly large $\lambda$ and small $\kappa\,\Delta\Theta^2$; and for longitudinal modes mostly large $\kappa\,\Delta\Theta^2$ -- see Fig.\,\ref{fig:M_Gamma_results}) are found where also the escape through the according channel is energetically favourable (see Fig.\,\ref{fig:comp_E}). In the disordered regime of mixed radial and longitudinal modes the escape is mostly inhibited and the type of the escape events occurring initially or at very large energies is less definite.

\section{Summary}

We have studied the dynamics of a two-dimensional ring of interacting units and its escape over the brim of a Mexican-hat-like potential under microcanonical conditions. We have identified and analysed radial breather modes and longitudinal resonant wave modes and where thus able to classify different parameter settings into three dynamical regimes. The system can either be dominated by one of the modes or will evolve towards a highly unordered state. This has a crucial impact on typical escape times. 

Escape can be realised via two escape channels which are related to different transition states. The escape through one of these channels is enhanced by breather modes. They efficiently accumulate energy into single radial degrees of freedom. This can cause single oscillators to pass over the potential's brim and subsequently pull the entire chain out of the meta-stable configuration. The second possibility is that the chain first overcomes the central potential hump and then surpasses the brim as a bundle. This escape path strongly relies on the presence of longitudinal resonant wave modes with wave number $m_{\varphi}=2$. In both of these cases early escapes occur already for energy values in the order of a few times the activation energy. Contrarily, the unordered dynamical regime practically prevents escape events even for significantly larger energies.

In a more general context this work shows that nonlinear cooperative effects among interacting units, which crucially impact the escape behaviour, are not an inherent property of highly idealized systems but remain relevant for complex potential landscapes as well.

\begin{acknowledgments}

L. Schimansky-Geier thanks for support from IRTG 1740 of the Deutsche Forschungsgemeinschaft.
\end{acknowledgments}
%%-Anhang----------------------------------------------------------
%
\appendix
\section{Stability of minimum energy configuration\label{App:stabl_fp}}
The full chain system as described by Eq. \eqref{qdyn1} has a fixed point, which is found for the coordinates (expressed in polar coordinates) $\varphi_i=i\,\Delta\Theta$, $r_i=r^0$, with $r^0$ defined in Eq. \eqref{r^0-conditional_equation} (minimum energy configuration), and momenta $\mathbf{p}_i=0$. Let us examine its stability through a linear stability analysis. We thus investigate the system for small displacements $\left| \boldsymbol\varepsilon_{i,1} \right|\ll 1$ of the coordinates from the minimum energy configuration and for small momenta $\left| \boldsymbol\varepsilon_{i,2} \right|\ll 1$. Neglecting all displacement terms of higher order, the linearised system can be written in block matrix form
  \begin{align}\label{lin_stability}  
  \begin{pmatrix}
  \dot{\varepsilon ^x}_{1,1}\\
  \dot{\varepsilon ^y}_{1,1}\\
  \dot{\varepsilon ^x}_{2,1}\\
  \vdots \\
  \dot{\varepsilon ^y}_{N,1}\\
  \dot{\varepsilon ^x}_{1,2}\\
    \vdots \\
  \dot{\varepsilon ^y}_{N,2} 
    \end{pmatrix}=
    \underbrace{
  \begin{pmatrix}
    0 		& \cdots  & 0	   &        &        &     \\
     \vdots & \ddots & \vdots &     & \mathbbold{1}_{2N} & \\
    0		& \cdots & 0 	  &     &     &  \\
            &        &        & 0 		& \cdots  & 0	  \\
            & J_{\boldsymbol f}      &        &  \vdots & \ddots & \vdots  \\
            &        &        & 0		& \cdots & 0 	 
  \end{pmatrix}}_{\displaystyle \equiv\mathbb{M} } 
   \begin{pmatrix}
 {\varepsilon ^x}_{1,1}\\
 \vdots\\
  {\varepsilon ^y}_{N,1}\\
  {\varepsilon ^x}_{1,2}\\  
   \vdots\\
     {\varepsilon ^y}_{N,2}\\  
    \end{pmatrix},
\end{align}
where the Jacobi matrix $J_{\boldsymbol f}$ is defined as the matrix of all first partial derivatives of the RHS of Eq. \eqref{qdyn1} with respect to the coordinates, which takes the form
\begin{align*}
\setlength{\arraycolsep}{2pt}
J_{\boldsymbol f}=
\begin{pmatrix}
\aleph_{x,1} & {\beth}_1 & \kappa & 0  & 0 & & \cdots & & \kappa & 0 \\
{\beth}_1 & \aleph_{y,1} & 0 & \kappa & 0  & & \cdots & & 0 & \kappa\\
\kappa & 0 & \aleph_{x,2} & {\beth}_2 & \kappa & 0 & 0 & \cdots & & 0 \\
0 & \kappa & {\beth}_2 & \aleph_{y,2} & 0 & \kappa & 0 & \cdots & & 0 \\
& & & & \vdots & & & & & \\
\kappa & 0 & 0 & & \cdots & & \kappa & 0 & \aleph_{x,N} & {\beth}_N \\
0 & \kappa & 0 & & \cdots & & 0		 & \kappa & {\beth}_N & \aleph_{x,N}
\end{pmatrix},
\end{align*}
with,
\begin{align*}
\aleph_{x,i}&=-2\,\kappa + \frac{{q^y_{i}}^0\,{q^y_{i}}^0}{{r^0}\,{r^0}\,{r^0}}\left(1+\frac{1}{\lambda}\sin\frac{r^0}{\lambda} \right) +\frac{{q^x_{i}}^0\,{q^x_{i}}^0}{\lambda^2\,{r^0}\,{r^0}}\cos\frac{r^0}{\lambda} \\
\aleph_{y,i}&=-2\,\kappa + \frac{{q^x_{i}}^0\,{q^x_{i}}^0}{{r^0}\,{r^0}\,{r^0}} \left(1+\frac{1}{\lambda}\sin\frac{r^0}{\lambda} \right) + \frac{{q^y_{i}}^0\,{q^y_{i}}^0}{\lambda ^2\,{r^0}\,{r^0}}\cos\frac{r^0}{\lambda}\\
\beth _i&=\frac{{q^x_{i}}^0\, {q^y_{i}}^0}{\lambda\, {r^0} {r^0} } \left[-\frac{1}{r^0} \left(  1 + \sin\frac{r^0}{\lambda}\right) +\frac{1}{\lambda}  \cos\frac{r^0}{\lambda} \right].
\end{align*}
The stability of equation (\ref{lin_stability}) can be inferred from the eigenvalues $\nu$ of $\mathbb{M}$. Its associated characteristic polynomial is defined as $ \det\left(\mathbb{M}-\nu\cdot\mathbbold{1}_{4N}\right)=0$. The evaluation of the determinant can be simplified due to the block matrix structure of $\mathbb{M}$. As $J_{\boldsymbol f}$ and $(\nu\cdot\mathbbold{1}_{2N})$ commute, we can reduce the problem to the eigenvalue problem for $J_{\boldsymbol f}$ (this can be seen from Leibniz formula for determinants) $\det(J_{\boldsymbol f}-\nu^2 \cdot\mathbbold{1}_{2N})=0$. $J_{\boldsymbol f}$ is symmetric and real. Therefore its eigenvalues $\nu^2$ are real as well. If any of them is positive the resulting pair of eigenvalues of $\mathbb{M}$, $\pm |\nu|$, includes a (real) positive eigenvalue and the fixed point is thus unstable. For the case $\nu^2<0$, the eigenvalues of $\mathbb{M}$ are purely imaginary and linear stability analysis fails to make predictions on the system's stability. However 
recalling that the system is conservative a stability statement can yet be made. The Jacobi matrix $J_{\boldsymbol f}$ equals the (negative of the) Hessian matrix of the potential energy function $U$ (external potential plus coupling energy) and thus the eigenvalues $\nu^2$ correspond to the negative curvature of the potential energy surface along the eigenvectors. When all $\nu^2$ are negative, the potential energy surface has a minimum at the minimum energy configuration. As the fixed point entails vanishing kinetic energy and energy is conserved all phase space trajectory in the vicinity of the fixed point are bound to the potential basin. Thus, in the case $\nu^2<0$ the fixed point is Lyapunov stable. The numerical solution of the eigenvalue problem of $J_{\boldsymbol f}$ yields Fig.\,\ref{fig:min_E_config_stability}.

\section{Time derivative of the total (angular) momentum for the longitudinal wave solution\label{A:total_momenta}}
The time derivative of the absolute value of the total angular momentum for $N$ unit mass particles with polar coordinates $r_i$ and $\varphi_i$, writes
$|\dot{\mathbf{L}}|=\sum_{i=0}^{N-1} r_i\left( 2 \dot{r}_i\,\dot{\varphi}_i+r_i\,\ddot{\varphi}_i \right)$. To evaluate this expression for the longitudinal wave solutions we have to take the continuum limit and apply the expression for the radial components expressed in Eq. \eqref{para_reso:radial_assumption} and the solutions of Eq. \eqref{para_reso:pde} for the angular components. According to the earlier definitions this yields $r_i(t) \rightarrow \widetilde{r}\,^0(t)$ and $\varphi_i(t)\rightarrow\varphi(\Theta,t) =\Phi(\Theta)\,T(t)$.
Thus the summation over all oscillators becomes an integral over $\Theta$, such that
\begin{align*}
|\dot{\mathbf{L}}|&=\widetilde{r}\,^0\left(2\,\dot{\widetilde{r}}\,^0\, \dot{T} \int_0^{2\pi}\Phi\,d\Theta+\widetilde{r}\,^0\, \ddot{T} \int_0^{2\pi}\Phi\,d\Theta \right),\\
\Rightarrow|\dot{\mathbf{L}}|&=0 \Leftrightarrow \int_0^{2\pi}\Phi\,d\Theta=0 \Leftrightarrow m_{\varphi}\neq 0.
\end{align*}

For the time derivative of the total momentum we find
\begin{align*}\dot{\mathbf{P}}=&\sum_{i=0}^{N-1}\left(\ddot{r}_i-r_i\,\dot{\varphi}_i^2\right)\begin{pmatrix}
\cos \varphi_i\\ \sin \varphi_i
\end{pmatrix}\\
+&\sum_{i=0}^{N-1}\left(2\,\dot{r}_i\,\dot{\varphi}_i+r_i\,\ddot{\varphi}_i\right)\begin{pmatrix}
-\sin \varphi_i\\
\cos \varphi_i
\end{pmatrix}.
\end{align*}
The conservation of the total momentum holds (up to corrections in the order of the initial random perturbations) as long as the chain resides in the $k=0$ phonon-like state, in which $\varphi_i= i\,\Delta\Theta$. After its initial preparation, the chain remains in such a setting until the onset of either radial or longitudinal wave modes. Thus, we can define a point in time $T_{\mathrm{init}} = \min\left(\Gamma_\varphi^{-1},\; \Gamma_r^{-1}\right) $ up until which the total momentum is conserved.

Taking the continuum limit and replacing the factors of the above equation by the according expressions in Eqs. \eqref{rdyn1} and \eqref{phidyn1}, the initial total angular momentum writes
\begin{gather*}\dot{\mathbf{P}}(t\ll T_{\mathrm{init}})=\int_0^{2\pi}d\Theta\,\left(1 + \frac{1}{\lambda}\sin\left(\frac{\widetilde{r}\,^0(t)}{\lambda}\right)\right)\begin{pmatrix}
\cos \Theta\\ \sin \Theta
\end{pmatrix}\\
+\int_0^{2\pi}d\Theta\,\left(-\kappa\left(\Delta\Theta\right)^2\widetilde{r}\,^0(t) \frac{\partial ^2 \varphi(\Theta,t)}{\partial \Theta ^2}\right)\begin{pmatrix}
-\sin \Theta\\
\cos \Theta
\end{pmatrix}\\
\propto  \int_0^{2\pi} d\Theta \left[\Phi_a^0\,\sin (m_{\varphi}\,\Theta) + \Phi_b^0\,\cos (m_{\varphi}\,\Theta)\right] \begin{pmatrix}
-\sin \Theta\\
\cos \Theta
\end{pmatrix}\\
\dot{\mathbf{P}}(t\ll T_{\mathrm{init}})=0 \Leftrightarrow m_{\varphi} \neq 1. 
\end{gather*}

\bibliography{my_bibliography}

\end{document}